\documentclass[11pt]{article}
\usepackage[utf8]{inputenc}
\usepackage{todonotes}
\usepackage{amsmath}
\usepackage{amssymb}
\usepackage{hyperref}
\usepackage{cleveref}
\usepackage{tikz}
\usepackage{subcaption}

\usepackage{authblk}
\usepackage{cite}
\usetikzlibrary{decorations.pathreplacing,calc,tikzmark}

\newlength{\abstractwidth}
\newcommand{\ads}{\mathrm{AdS}}
\newcommand{\Gbulk}{G_\mathrm{N}}
\newcommand{\Gbrane}{G_\mathrm{brane}}

\newcommand{\brn}{\mathrm{brane}}

\newcommand{\lblk}{L}

\DeclareMathOperator{\tr}{\text{tr}}

\title{Homology Conditions for RT Surfaces in Double Holography}
\author{Dominik Neuenfeld}

\begin{document}

\begin{titlepage}
\begin{center}
{\LARGE \bf
Homology Conditions for RT Surfaces in Double Holography

}
\vspace{1cm}

\textbf{Dominik Neuenfeld}
\vspace{0.5cm}

\textit{Perimeter Institute for Theoretical Physics, Waterloo, ON N2L 2Y5, Canada }
\vspace{0.5cm}
 
\texttt{dneuenfeld@pitp.ca}
\end{center}

\vspace{2cm}
\begin{abstract}Recently, a novel formula for computing entropy in theories coupled to semi-classical gravity has been devised. Using this so-called island formula the entropy of semi-classical black holes follows a Page curve. Here, we study the relation between this novel entropy and semi-classical entropy in the context of doubly-holographic models.

Double holography allows for two different $d$-dimensional descriptions of a black hole coupled to a non-gravitational bath, both of which allow a holographic computation of von Neumann entropy in bath subregions. We argue that the correct homology constraint for Ryu-Takayanagi surfaces depends on which of those $d$-dimensional perspectives is taken. As a consequence the von Neumann entropies of a fixed subregion in both descriptions can disagree. We discuss how the von Neumann entropies in both descriptions are related to the entropy computed by the island formula and coarse grained entropy. Moreover, we argue that the way operators transform between the two descriptions depends on their complexity.
A simple toy model is introduced to demonstrate that a sufficiently complicated map between two descriptions of the system can give rise to an island formula and wormholes. Lastly, we speculate about the relation between double-holography and black hole complementarity.
\end{abstract}

\end{titlepage}

\clearpage

\section{Introduction}

\subsection{Islands in Semi-Classical Gravity}
In the last few years it was discovered how to compute the Page curve \cite{Page:1993wv} from semi-classical gravity \cite{Penington:2019npb, Almheiri:2019psf, Almheiri:2019hni} in a controlled manner. Instead of considering the von Neumann entropy
\begin{align}
    \label{eq:vN_intro}
    S_\text{vN}(A) = - \tr \rho_A \log \rho_A
\end{align}
of a collection of Hawking quanta \cite{Hawking:1974rv,Hawking:1974sw} described by a reduced density matrix $\rho_A$ and contained in a region $A$ outside the black hole, one can obtain the Page curve if one instead uses the quantum extremal surface prescription \cite{Engelhardt:2014gca} to compute the entropy \cite{Penington:2019npb, Almheiri:2019psf, Almheiri:2019hni},
\begin{align}
    \label{eq:island_intro}
    S_\text{island}(A) = \underset{B}{\min \text{ext}} \left\{ \frac{\text{Area}(\partial B)}{4 \Gbulk} + S_\text{vN}(A \cup B)\right\}.
\end{align}
This formula has been called the island formula in \cite{Almheiri:2019hni} where it has been suggested that it computes the ``fine-grained'' or ``quantum gravitational'' entropy of radiation contained in a subregion $A$. For a recent review, see \cite{Almheiri:2020cfm}. In this paper we will call the quantity computed by \cref{eq:island_intro} \emph{island entropy} in order to distinguish it from other measures of entropy, such as \cref{eq:vN_intro}.

The island formula has been derived in two-dimensional JT gravity coupled to a CFT using Euclidean replica wormholes \cite{Almheiri:2019qdq}, see also \cite{Penington:2019kki}. It has been applied to a variety of different situations in two or higher dimensions where again it has been shown to evade information paradoxes, e.g., \cite{Saad:2019lba,Saad:2019pqd,Saad:2021rcu, Gautason:2020tmk, Hashimoto:2020cas, Hartman:2020swn, Krishnan:2020oun, Alishahiha:2020qza, Balasubramanian:2020xqf, Chen:2020jvn, Chen:2020tes, Rozali:2019day, Almheiri:2019psy, Geng:2020qvw, Chen:2020uac, Chen:2020hmv, Hernandez:2020nem, Bhattacharya:2021jrn}. Apart from the two-dimensional case, the island formula has also been argued for in higher dimensions in semi-classical gravity \cite{Marolf:2020rpm}. Moreover, in higher dimensions it has been derived in the context of doubly-holographic models \cite{Rozali:2019day, Almheiri:2019psy, Geng:2020qvw, Chen:2020uac, Chen:2020hmv, Hernandez:2020nem}. In this latter case, the island formula arises as a consequence of the standard Ryu-Takayanagi \cite{Ryu:2006bv, Ryu:2006ef, Hubeny:2007xt} prescription.

Since the standard RT prescription as well as holography are well-understood, this makes doubly-holographic models an ideal playground to investigate the origin and interpretation of the island formula. And in fact, despite much progress, there are still many important questions that have not been conclusively answered. What is the microscopic origin of the island formula in higher dimensions \cite{Bousso:2019ykv, Pollack:2020gfa, Marolf:2020xie, Belin:2020hea, Bousso:2020kmy, Stanford:2020wkf}? What is the precise relation between the fine-grained description of gravity which gives rise to \cref{eq:island_intro}, and semi-classical gravity, where entropies are naively computed using \cref{eq:vN_intro} \cite{Harlow:2020bee}? What was the mistake in Hawking's calculation \cite{Hawking:1976ra, Hawking:1976de}?

Doubly-holographic models are constructed by introducing an end-of-the-world (ETW) brane,\footnote{These branes are also called \emph{Planck branes} or \emph{Karch-Randall branes}.} whose extrinsic curvature has constant trace, into an asymptotically $\ads_{d+1}$ space \cite{Randall:1999vf, Karch:2000ct}. This description of the system is referred to as the \emph{bulk perspective}. Via the AdS/BCFT correspondence \cite{Maldacena:1997re, Witten:1998qj, Gubser:1998bc, Aharony:1999ti, Takayanagi:2011zk, Fujita:2011fp} this system can be described by a BCFT$_d$ with degrees of freedom on its $(d-1)$-dimensional boundary \cite{Takayanagi:2011zk, Fujita:2011fp}.\footnote{There are of course other types of doubly-holographic scenarios where the ETW brane is either replaced by a defect in the bulk and the dual is a defect CFT, or the ETW brane intersects the boundary in Euclidean time and the dual is a particular CFT microstate. We will however focus on the case where the boundary theory is a BCFT.} We will call this BCFT description the \emph{boundary perspective} and as usual assume that it serves as the definition of quantum gravity in the bulk. In the limit of a large number of boundary degrees of freedom the system exhibits a third, $d$-dimensional, approximate description as a CFT$_d$ coupled to semi-classical gravity, called the \emph{brane perspective}. CFT excitations can move between the rigid background spacetime of the ambient BCFT and the dynamical spacetime of the ETW brane, whose dynamics is controlled by Einstein gravity plus higher order corrections\footnote{In the model considered here, the graviton has a small mass \cite{Karch:2000ct, Schwartz:2000ip, Miemiec:2000eq, Porrati:2001db, Neuenfeld:2021wbl, Geng:2020fxl}.}. For concreteness, we will be interested in a doubly-holographic model describing a $d$-di\-men\-si\-o\-nal eternal topological black hole coupled to and in equilibrium with a non-gravitating bath\footnote{For discussions of gravitating baths, see \cite{Geng:2020fxl, Ghosh:2021axl, Anderson:2021vof, Geng:2021iyq, Balasubramanian:2021wgd}.} \cite{Chen:2020uac,Chen:2020hmv}. It has been shown in \cite{Chen:2020uac} that in this model the island formula in the brane perspective is nothing but the standard RT formula for the BCFT description.

\subsection{Summary of Results}
As we will explain in this paper, in doubly-holographic models the island formula does \emph{not} compute the von Neumann entropy of Hawking radiation of a given bath subregion $A$ in the brane perspective. Rather, it computes the entropy of radiation in the same region $A$ in the boundary perspective. It thus is a proxy for computing the von Neumann entropy of $A$ in a different formulation of the theory, which we can think of as the true, quantum gravitational description. Instead of a black hole, this formulation contains a defect with a large number of degrees of freedom. In this paper, assuming that standard facts about entanglement wedge reconstruction are correct in both the brane and boundary perspective, we will argue that for a bath subregion $A$ in double holography the following relations hold:
\begin{align}
\label{eq:result_entropies}
\begin{split}
S_\text{island}(A) \text{ in the \textbf{brane} perspective} &= \;\;\;\; S_\text{vN}(A) \text{ in the \textbf{boundary} perspective,}\\
S_\text{vN}(A) \text{ in the \textbf{brane} perspective} &= S_\text{simple}(A) \text{ in the \textbf{boundary} perspective.}
\end{split}
\end{align}
Here, $S_\text{simple}(A)$ is the simple entropy introduced in \cite{Engelhardt:2017aux, Engelhardt:2018kcs} and recently refined in \cite{Engelhardt:2021mue, Engelhardt:toappear}. The simple entropy of a state $\rho$ is a coarse grained entropy which measures the number of states $\rho'$ which give the same one-point functions for simple operators in the presence of simple sources than the original state $\rho$. The definition of simple operators and simple sources can be found in \cite{Engelhardt:2017aux, Engelhardt:2018kcs, Engelhardt:2021mue}, but will also be reviewed in \cref{sec:mapping}.

This prescription has a puzzling consequence whose resolution lies at the heart of this paper: The von Neumann entropy of a fixed region $A$ depends on a choice of perspective. 
In the bulk this can be understood through the RT formula. Whether we compute $S_\text{vN}(A)$ in the boundary or the brane perspective is reflected by the choice of a homology constraint for the bulk RT surface. If one wants to compute the von Neumann entropy in the boundary perspective the RT surface must be homologous to the boundary region $A$ up to subsets which lie within the ETW brane \cite{Takayanagi:2011zk, Fujita:2011fp}. This becomes particularly clear in the model of \cite{Chen:2020uac, Chen:2020hmv, Hernandez:2020nem}. If we want to compute the von Neumann entropy in the brane perspective, the usual homology constraint applies, namely, the RT surface can only end on the ETW brane if the intersection is part of the boundary of $A$.

From the brane and boundary perspective, this puzzle is resolved by entanglement wedge reconstruction. Assuming that entanglement wedge reconstruction works as usual, i.e., given a subregion $A$ we can reconstruct the bulk within the entanglement wedge associated to $A$, it is a straightforward consequence that the same part of the bulk can get encoded into different degrees of freedom in either of the dual perspectives. As it turns out, simple operators in the bath agree in the boundary and brane perspective. However, certain complex operators in the boundary perspective are non-locally related to simple operators in the brane perspective.

Under certain assumptions on entanglement wedge reconstruction, we will show that the resulting picture in double holography is as follows: There are two $d$-dimensional formulations for an eternal black hole in equilibrium with radiation. One formulation, the boundary perspective, describes the region around and including the black hole as a quantum mechanical defect. Here, the black hole is a quantum mechanical system, but violates the equivalence principle, since in this description spacetime ends on the defect. Information about what fell into the black hole is available from the Hawking radiation after the Page time, although it will be difficult to access. The other formulation, the brane perspective, models the black hole by quantum field theory on a semi-classical background. Here, the equivalence principle holds, but information never leaves the black hole. Thus, the information about the black hole interior exists only once in either formulation. In particular this means that unitarity for an outside observer is violated. Still, simple observables (such as low point correlation functions) agree with simple observables in the boundary perspective. While for finite time and simple operations outside the black hole both pictures agree, they significantly differ if one allows complex operations, such as measuring the S-matrix or determining whether the Hawking radiation is pure. However, even though the boundary perspective should be used to describe acting with complex operators, there are situations in which complex quantities in the boundary perspective can be translated to easy-to-compute (but non-local) quantities in the brane perspective---one example being the island formula. In such a situation, an observer which can perform complex operations on the Hawking radiation in the boundary perspective seems to have non-local control over the black hole interior in the brane perspective.

Lastly, we speculate that the structure exhibited by the doubly-holographic model is a viable mechanism for black hole complementarity. That is, there are (at least) two complementary formulations of a black hole, one of which is compatible with the equivalence principle but does not return information from the black hole, and another one which returns information from a black hole but does not agree with the equivalence principle. An example of the former perspective is semi-classical gravity, while an example of the latter formulation would be fuzzballs \cite{Mathur:2005zp} or a stretched horizon \cite{Susskind:1993if}. If true, this would mean that one does not need to rely on physical intuition to argue for how complementarity might work, instead we can study double holography.

\subsection{Structure of the Paper}
The rest of this paper is structured as follows. We start \cref{sec:rt_surfaces} by briefly reviewing the doubly-holographic model we are interested in, before arguing for our main claim in \cref{sec:rt_surfaces_in_brane}, namely that the homology constraint for bulk RT surfaces should depend on whether we want to compute von Neumann entropies in the boundary or brane perspective. The next section, \cref{sec:mapping}, elaborates on the consequences for the relation between entropies and entanglement wedge reconstruction, and how a number of quantities are mapped between the brane and the boundary description. In \cref{sec:model} we introduce a crude toy model which illustrates how a sufficiently complicated mapping between Hilbert spaces can give rise to the island formula as well as wormholes. \Cref{sec:bhc} is more speculative. We point out that a version of black hole complementarity might be realized which is very similar to the duality between the boundary and brane perspective, and start exploring the consequences of this observation. In \cref{sec:discussion} we conclude by pointing out unexplored connections between our and existing work as well as new directions.

\section{Homology Constraints for RT Surfaces in Double Holography}
\label{sec:rt_surfaces}
We will start this section by establishing the doubly-holographic model we will be working with, including a brief review of the different layers of holography: The bulk perspective, the brane perspective, and the boundary perspective. We will not discuss technical details for which we refer the reader to the existing literature \cite{Karch:2000ct, Aharony:2003qf, Almheiri:2019hni, Chen:2020uac, Chen:2020hmv}. The remainder of this section then discusses how the homology constraint for RT surfaces in the bulk changes depending on whether we are computing von Neumann entropies for the boundary or brane perspective.

\subsection{An Eternal Black Hole in Double Holography}
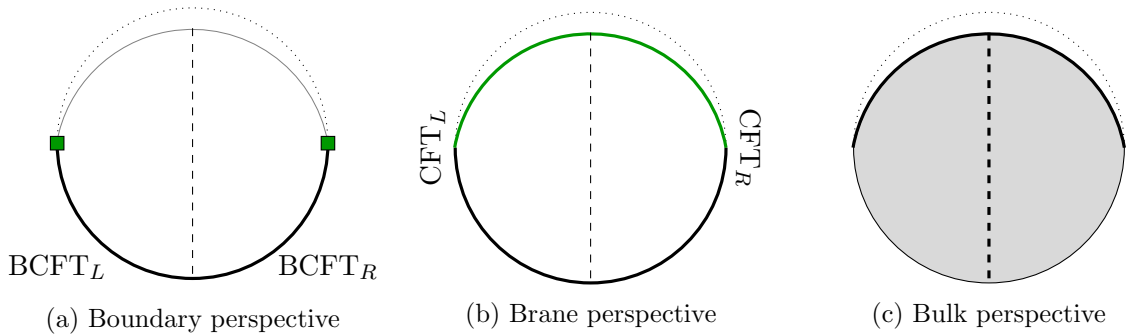
\begin{figure}[t]
    \centering
    \begin{subfigure}{.32\linewidth}
    \centering
    \begin{tikzpicture}[scale=0.9]
    \draw[black, very thick] (0,0) arc (0:-180:2cm);
    \draw[dotted] (0,0) arc (0:180:2cm); 
    \draw[very thin, gray] (0,0) arc (10:170:2.035cm);
    \fill[black!40!green, draw=black] (-0.1,-0.1) rectangle (0.1,0.1);
    \fill[black!40!green, draw=black] (-4.1,-0.1) rectangle (-3.9,0.1);
    \draw[dashed] (-2,1.7) -- (-2, -2);
    \node[below] at (0, -1.5) {BCFT$_R$};
    \node[below] at (-4, -1.5) {BCFT$_L$};
    \end{tikzpicture}
    \caption{Boundary perspective}
    \label{fig:boundary_perspective}
    \end{subfigure}
    \begin{subfigure}{.32\linewidth}
    \centering
    \begin{tikzpicture}[scale=0.9]
    \draw[black, very thick] (0,0) arc (0:-180:2cm);
    \draw[dotted] (0,0) arc (0:180:2cm); 
    \draw[black!40!green, very thick] (0,0) arc (10:170:2.035cm);
    \draw[dashed] (-2,1.7) -- (-2, -2);
    \node[above, rotate=270] at (0, 0) {CFT$_R$};
    \node[above,rotate=90] at (-4, 0) {CFT$_L$};
    \end{tikzpicture}
    \caption{Brane perspective}
    \label{fig:brane_perspective}
    \end{subfigure}
    \begin{subfigure}{.32\linewidth}
    \centering
    \begin{tikzpicture}[scale=0.9]
    \fill[white!70!gray, draw=black] (0,0) arc (0:-180:2cm) arc (170:10:2.035cm);
    \draw[black, very thick] (0,0) arc (10:170:2.035cm);
    \draw[dotted] (0,0) arc (0:180:2cm); 
    \draw[dashed, very thick, draw=black] (-2,1.7) -- (-2, -2.);
    \end{tikzpicture}
    \caption{Bulk perspective}
    \label{fig:bulk_perspective}
    \end{subfigure}
    \caption{Time slices of the the three layers of holography in a doubly-holographic setting with an eternal topological black hole. \textbf{(a)} The boundary perspective consists of two entangled BCFTs coupled to lower-dimensional theories at their respective boundaries (green squares). \textbf{(b)} In the brane perspective the boundary degrees of freedom get replaced by a gravitating region (green line) which hosts a black hole. The horizon is located at the intersection of the brane with the bulk horizon (dashed line). \textbf{(c)} The bulk perspective describes the system as a two-sided, eternal, massless topological AdS black hole. The bifurcate horizon is indicated by the black dashed line. The asymptotic boundary of the upper part of the bulk geometry (dotted line) is cut off by an ETW brane (thick black line).}
    \label{fig:three_perspectives}
\end{figure}
We will consider a doubly-holographic model which describes an eternal, massless, topological black hole coupled to a bath at the same temperature. Such a model was introduced in two dimensions in \cite{Almheiri:2019hni} and in higher dimensions in \cite{Chen:2020hmv}\footnote{More precisely we discuss a $\mathbb Z_2$ quotient of the model introduced in \cite{Chen:2020hmv}.}, see \cref{fig:three_perspectives}. The bulk geometry is given by vacuum $\ads_{d+1}$ viewed in AdS Rindler coordinates. Part of the $\ads$ boundary is cut off by a ETW brane which perpendicularly intersects the bulk Rindler horizon. The action is schematically given by
\begin{align}
    S = S_\text{bulk} + S_\text{GHY} + S_\brn,
\end{align}
where $S_\text{bulk}$ is the bulk action, $S_\text{GHY}$ is the Gibbons-Hawking-York boundary term at the asymptotic boundary as well as the brane, and the brane action is given by
\begin{align}
    S_{\brn} = - T_0 \int_\brn \sqrt{- \gamma}.
\end{align}
Neumann boundary conditions at the brane relate its location to its tension $T_0$. Generally, one might want to add local terms of the induced metric $\gamma_{ij}$ and other fields to the brane action \cite{Dvali:2000hr}. This becomes particularly important in two brane dimensions if one wants to recover JT gravity in the brane perspective \cite{Almheiri:2019hni, Chen:2020uac}. A zero tension brane would intersect the bulk geometry at a $\mathbb Z_2$ symmetric surface so that we are only left with the lower half of the time slice in \cref{fig:bulk_perspective}. As we increase the tension, more and more spacetime is created by the backreaction of the brane. Equivalently, we can say that the brane moves outwards in the direction where the cut-off part of the asymptotic boundary would be. The brane tension has to be smaller than a critical value\footnote{A brane whose tension equals the critical tension would naively sit exactly at the asymptotic boundary. The correct model for this case is the Randall-Sundrum model \cite{Randall:1999vf}.},
\begin{align}
    \label{eq:critical_tension}
    T_0 < T_\text{crit} = \frac{d-1}{8\pi \Gbulk \lblk}.
\end{align}
We will call this bulk representation of our spacetime the \emph{bulk perspective}.

To each side of the horizon, the dual description---called \emph{boundary perspective}---is given by a BCFT coupled to a large number of additional degrees of freedom at its boundary, \cref{fig:boundary_perspective}. The state of the system has been prepared by a Eucliden path integral such that at $t=0$ both BCFTs are in the thermofield double state and live on a hyperbolic background with curvature radius $R$. The whole system is at a finite temperature given in terms of the background curvature scale by
\begin{align}
    T = \frac 1 {2 \pi R}.
\end{align}
In the boundary perspective, we will refer to the $d$-dimensional ambient spacetime, i.e., the region away from the boundary degrees of freedom, as the \emph{bath}. Correlation functions in the BCFT can be computed by the usual methods of holographic renormalization. For further technical details see \cite{DeWolfe:2001pq, Takayanagi:2011zk, Fujita:2011fp}. Like in AdS/CFT, this formulation serves as the definition of quantum gravity in the bulk perspective.

As the brane tension approaches the critical value, \cref{eq:critical_tension}, a third description of the system becomes useful \cite{Randall:1999vf, Karch:2000ct}: the \emph{brane perspective}, \cref{fig:brane_perspective}. Taking this perspective, the theory on each side of the horizon is described by an effective CFT which lives on a background consisting of two pieces glued together. The first piece of the geometry is the ambient, rigid spacetime of the BCFT in the boundary perspective---we will again call this the \emph{bath}---while the second piece is a gravitating, eternal $\ads$ black hole geometry. The latter part corresponds to the geometry of the brane on which the bulk metric induces a topological black hole. The geometry in this region is dynamical and controlled by Einstein-Hilbert gravity plus higher curvature corrections. The matter theory on the brane is the same as the bath CFT. Excitations are free to cross from the non-gravitating bath region into the gravitating region and fall into the black hole horizon. However, in this formulation the CFT on the brane is only an effective theory. The description comes with a cutoff set by the location of the brane
\begin{align}
    \Lambda_\text{cutoff} = \lblk, && \ell = \frac{L}{\sqrt{1 - \left(\frac{T_0}{T_\text{crit}}\right)^2}}.
\end{align}
Here $\lblk$ is the bulk AdS scale, while $\ell$ is the brane AdS scale. The description in terms of a local CFT coupled to gravity on a semi-classical background is good if $\ell \ll L$. Again, correlation functions in this perspective can be computed from bulk data \cite{Gubser:1999vj, PerezVictoria:2001pa, Porrati:2001gx, Aharony:2003qf, Neuenfeld:2021wbl}. For energies of the order $\Lambda_\text{cutoff}$ the higher dimensional nature of the theory becomes apparent. For this reason, we should only trust the brane perspective if excitations on the brane only cause small backreaction. We now want to understand how to compute entropies in either the brane or the boundary perspective through a bulk calculation.

\subsection{RT Surfaces and Entropies in the Boundary Perspective}
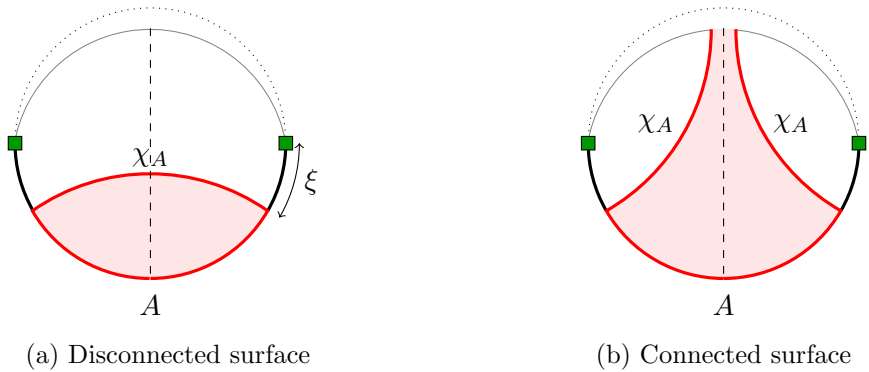
\begin{figure}[t]
    \centering
    \begin{subfigure}{.45\linewidth}
    \centering
    \begin{tikzpicture}[scale=0.9]
    \draw[dotted] (0,0) arc (0:180:2cm); 
    \draw[very thin, gray] (0,0) arc (10:170:2.035cm);
    \draw[black, very thick] (0,0) arc (0:-30:2cm);
    \draw[black, very thick] (-4,0) arc (-180:-150:2cm);
    \fill[black!40!green, draw=black] (-0.1,-0.1) rectangle (0.1,0.1);
    \fill[black!40!green, draw=black] (-4.1,-0.1) rectangle (-3.9,0.1);
    \fill[white!90!red, draw=red, very thick] (-2,-2) arc (-90:-30:2cm) arc (55:125:3.03cm) arc (-150:-90:2cm);
    \draw[dashed] (-2,1.7) -- (-2, -2);
    \node[below] at (-2, -2.1) {$A$};
    \node[above] at (-2, -0.5) {$\chi_A$};
    \draw[<->] (0.2,0) arc (0:-30:2.2cm) node[midway, right] {$\xi$};
    \end{tikzpicture}
    \caption{Disconnected surface}
    \label{fig:disconnected_RT_surface}
    \end{subfigure}
    \begin{subfigure}{.45\linewidth}
    \centering
    \begin{tikzpicture}[scale=0.9]
    \draw[black, very thick] (0,0) arc (0:-30:2cm);
    \draw[black, very thick] (-4,0) arc (-180:-150:2cm);
    \draw[dotted] (0,0) arc (0:180:2cm); 
    \draw[very thin, gray] (0,0) arc (10:170:2.035cm);
    \fill[black!40!green, draw=black] (-0.1,-0.1) rectangle (0.1,0.1);
    \fill[black!40!green, draw=black] (-4.1,-0.1) rectangle (-3.9,0.1);
    \fill[white!90!red, very thick] (-2,-2) arc (-90:-30:2cm) arc (-120:-180:3.1cm) arc (85:95:2.035) arc (0:-60:3.1cm) arc (-150:-90:2cm);
    \draw[red, very thick] (-2,-2) arc (-90:-30:2cm) arc (-120:-180:3.1cm);
    \draw[red, very thick] (-2,-2) arc (-90:-150:2cm) arc (-60:0:3.1cm);
    \draw[dashed] (-2,1.7) -- (-2, -2);
    \node[below] at (-2, -2.1) {$A$};
    \node[above] at (-3, 0) {$\chi_A$};
    \node[above] at (-1, 0) {$\chi_A$};
    \end{tikzpicture}
    \caption{Connected surface}
    \label{fig:connected_RT_surface}
    \end{subfigure}
    \caption{To compute the entanglement entropy of a boundary perspective subregion $A$ (red interval on the asymptotic boundary) we have to pick the smaller of two extremal candidate surfaces (dark red lines in the bulk). The region between the RT surface and the boundary is the entanglement wedge (pink shaded region). The \emph{disconnected} surface agrees with the RT surface one would expect in usual AdS/CFT. The \emph{connected} surface connects to the brane and has a much large entanglement wedge which reaches the brane.}
    \label{fig:RT_surfaces_boundary_perspective}
\end{figure}
The boundary perspective is equivalent to the description of quantum gravity on asymptotically $\ads$ space via the standard $\ads/$CFT correspondence in that the boundary serves as the UV complete definition of gravity in the bulk. As in usual AdS/CFT, the von Neumann entropy of the reduced density matrix of a boundary subregion $A$ can be computed in the bulk from the area of the associated Ryu-Takayanagi surface $\chi_A$. A new aspect of holographic BCFTs is that the RT surface is allowed to end on the bulk ETW brane \cite{Takayanagi:2011zk, Fujita:2011fp}.

The RT surface $\chi_A$ associated with the boundary region $A$ in a BCFT is defined to be the smallest codimension-two surface with extremal area\footnote{To be more precise, it must extremize the generalized area functional \cite{Engelhardt:2014gca}. However, for our purposes the quantum corrections to the area functional in the bulk are always small and can be ignored.}, homologous to $A$, up to terms on the ETW brane. That is, there exists a partial Cauchy surface $\Sigma_A$, such that
\begin{align}
    \label{eq:homology_bcft}
        \partial \Sigma_A = A \cup \chi_A \cup X, && X \subset \text{ETW brane}.
\end{align}
It will be sufficient for us to focus on subregions which consist of all points further than a distance $\xi$ away from the BCFT boundary, see \cref{fig:disconnected_RT_surface}. Given the RT surface we can compute $S_\text{vN}(A)$, defined via
\begin{align}
    S_\text{vN}(A) = - \tr \rho_A \log \rho_A, && \text{with } \rho_A = \tr_{\bar A} \rho,
\end{align}
by computing
\begin{align}
    S_\text{vN}(A) = \frac{\text{Area}(\chi_A)}{4 \Gbulk}.
\end{align}
As is well known, the definition of $S_\text{vN}(A)$ in the boundary theory is plagued by UV divergences which correspond to IR divergences of the area of the RT surface. Here, we will always implicitly use a regulator.

For a region $A$ away from the boundary defect there are two extremal candidate surfaces, see \cref{fig:RT_surfaces_boundary_perspective}. The first configuration, \cref{fig:disconnected_RT_surface}, is an extremal surface which caps off $A$. Since RT surfaces can end on the ETW brane, there is another possible configuration where the consists of two separate parts and connects to a location on the ETW brane, see \cref{fig:connected_RT_surface}.

The precise geometry of such surfaces has been discussed in \cite{Almheiri:2019hni, Rozali:2019day, Chen:2020uac, Chen:2020hmv}. There, it has also been shown that at early times (i.e., $|t|$ is small) the first configuration, \cref{fig:disconnected_RT_surface}, has smaller area and is thus the correct RT surface. As time goes on, the area of the RT surface grows linearly with time. This is because the RT surface falls into the black hole horizon and exits on the other side. Since the Einstein-Rosen bridge that connects the two sides grows in time, so does the RT surface \cite{Hartman:2013qma}. From the boundary point of view, the growing entanglement entropy signals information exchange between the region $A$ and the complement $\bar A$ which includes the boundary degrees of freedom of the BCFT.

After some time, the entanglement between the region $A$ and its complement $\bar A$ takes on its maximal value and the linear growth stops. In the boundary perspective this must happen since the BCFT boundary degrees of freedom only carry a finite density of degrees of freedom.  From a bulk point of view this happens because of a phase transition in the RT surface. The area of the first candidate surface, \cref{fig:disconnected_RT_surface}, exceeds that of the second surface, \cref{fig:connected_RT_surface}, which is constant in time and thus becomes the true RT surface. 

We will refer to the time at which the linear growth stops as the Page time $t_P$. This is motivated by the fact that in the brane perspective, where the system is described as a semi-classical black hole in contact with a bath, the model exhibits an information paradox \cite{Almheiri:2019yqk, Rozali:2019day}. From the brane perspective, the change in RT surface is interpreted as the appearance of an island outside the horizon of the black hole \cite{Almheiri:2019hni,Chen:2020uac,Chen:2020hmv}. Thanks to this, the entropy stops growing.

\subsection{RT Surfaces and Entropies in the Brane Perspective}
\label{sec:rt_surfaces_in_brane}
\begin{figure}[t]
    \centering
    \begin{subfigure}{.45\linewidth}
    \centering
    \begin{tikzpicture}[scale=0.9]
    \draw[black, very thick] (0,0) arc (0:-180:2cm);
    \draw[dotted] (0,0) arc (0:180:2cm); 
    \draw[black!40!green, very thick] (0,0) arc (10:170:2.035cm);
    \fill[white!90!red, draw=red, very thick] (-2,1.7) arc (90:50:2.035cm) arc (-30:-150:1.5cm) arc (130:90:2.035cm);
    \draw[dashed] (-2,1.7) -- (-2, -2);
    \node[below] at (-2, -2.1) {$A = \emptyset$};
    \node[above] at (-2, 1.7) {$B$};
    \node[below] at (-2, 0.4) {$\chi_B$};
    \end{tikzpicture}
    \caption{RT surface associated to $B$}
    \label{fig:RT_surface_B}
    \end{subfigure}
    \begin{subfigure}{.45\linewidth}
    \centering
    \begin{tikzpicture}[scale=0.9]
    \draw[black, very thick] (0,0) arc (0:-180:2cm);
    \draw[dotted] (0,0) arc (0:180:2cm); 
    \fill[white!90!red, draw=red, very thick] (0,0) arc (10:170:2.035cm) -- (0,0);
    \draw[dashed] (-2,1.7) -- (-2, -2);
    \node[below] at (-2, -2.1) {$A = \emptyset$};
    \node[above] at (-2, 1.7) {$B$};
    \node[below] at (-2, -0.1) {$\chi_B$};
    \end{tikzpicture}
    \caption{RT surface associated to larger region $B$}
    \label{fig:RT_surface_B_big}
    \end{subfigure}
    \caption{The generalized entropy of a subregion of the gravitation region can be computed using the RT formula. This implies that the bulk region close to the brane is always in the entanglement wedge of the gravitational region---even after the Page time.}
    \label{fig:RT_gravity_subregion}
\end{figure}
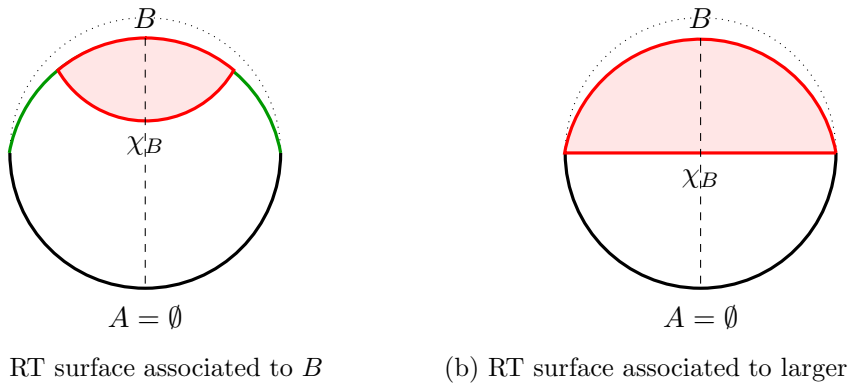
The brane perspective is merely an effective description of the system at $T_\text{crit} - T_0 \ll 1$. Nonetheless, we can still compute entropies using the RT prescription. In fact, one might hope that this is even possible for subregions on the brane, i.e., where we have dynamical gravity\footnote{Note that in a dynamical theory of gravity it is a priori not clear how subregions can be defined. Here, we have various options. For example, we could send light rays from the boundary and define a subregion by requiring that a mirror placed at the boundary of this subregion reflects the light such that it intersects the boundary a second time at a given location. However, there is an even easier way. Since we anyways only trust the brane perspective if there is small backreaction on the brane, we can fix our gauge and define a region in gauge-fixed background coordinates.}.
There are several non-trivial arguments that suggest this is indeed the case. As was shown in \cite{Chen:2020uac}, the leading order contribution to RT surfaces ending on $\partial B$ on the brane can be matched with terms of the Wald-Dong entropy \cite{Wald:1993nt,Dong:2013qoa} of $B$. It was shown that in the brane perspective the entropy computed using the RT surface in \cref{fig:connected_RT_surface} can be written as 
\begin{align}
    S_\text{RT} = \frac{A(\partial B)}{4 \Gbrane} + S(A \cup B) + \text{(higher curvature corrections)},
\end{align}
where $A$ is a bath subregion and $B$ is the \emph{island}, a non-empty subset of the gravitational region which extremizes \cref{eq:island_intro}. This supports the claim that RT surfaces ending on the boundary of some gravitating subregion $B$ computes the generalized entropy of $B$. This observation should not be very surprising, since as we get close to the critical limit of brane tension, the brane essentially acts like an IR regulator in standard AdS/CFT. It thus makes sense to treat the theory on the brane following the same rules as the CFT in the standard AdS/CFT correspondence, including the computation of RT surfaces. Lastly, recent work \cite{Geng:2021wcq, Ghosh:2021axl} also indicates that this interpretation should be taken to correctly reproduce the Page curve in settings where the bath is gravitating.

\begin{figure}[t]
    \centering
    \begin{subfigure}{.45\linewidth}
    \centering
    \begin{tikzpicture}[scale=0.9]
    \draw[black, very thick] (0,0) arc (0:-30:2cm);
    \draw[black, very thick] (-4,0) arc (-180:-150:2cm);
    \draw[dotted] (0,0) arc (0:180:2cm); 
    \draw[black!40!green, very thick] (0,0) arc (10:170:2.035cm);
    \fill[white!90!red, very thick, draw = red] (-2,-2) arc (-90:-30:2cm) arc (-120:-180:3.1cm) arc (85:95:2.035) arc (0:-60:3.1cm) arc (-150:-90:2cm);
    \draw[dashed] (-2,1.7) -- (-2, -2);
    \node[below] at (-2, -2.1) {$A$};
    \node[above] at (-3.2, 0) {$\chi_{A \cup B}$};
    \node[above] at (-0.8, 0) {$\chi_{A \cup B}$};
    \node[above] at (-2, 1.7) {$B \neq \emptyset$};
    \end{tikzpicture}
    \caption{Allowed configuration}
    \label{fig:RT_surface_allowed}
    \end{subfigure}
    \begin{subfigure}{.45\linewidth}
    \centering
    \begin{tikzpicture}[scale=0.9]
    \draw[black, very thick] (0,0) arc (0:-30:2cm);
    \draw[black, very thick] (-4,0) arc (-180:-150:2cm);
    \draw[dotted] (0,0) arc (0:180:2cm); 
    \fill[white!90!red, very thick] (-2,-2) arc (-90:-30:2cm) arc (-120:-180:3.1cm) arc (85:95:2.035) arc (0:-60:3.1cm) arc (-150:-90:2cm);
    \draw[red, very thick] (-2,-2) arc (-90:-30:2cm) arc (-120:-180:3.1cm);
    \draw[red, very thick] (-2,-2) arc (-90:-150:2cm) arc (-60:0:3.1cm);
    \draw[dashed] (-2,1.7) -- (-2, -2);
    \draw[black!40!green, very thick] (0,0) arc (10:170:2.035cm);
    \node[below] at (-2, -2.1) {$A$};
    \node[above] at (-3, 0) {$\chi_A$};
    \node[above] at (-1, 0) {$\chi_A$};
    \begin{scope}[transparency group,opacity=.6]
    \draw[line width = 0.5cm, gray] (-4,-2) -- (0,2);
    \draw[line width = 0.5cm, gray] (-4,2) -- (0,-2);
    \end{scope}
    \node[above] at (-2, 1.7) {$B = \emptyset$};
    \end{tikzpicture}
    \caption{Disallowed configuration}
    \label{fig:RT_surface_disallowed}
    \end{subfigure}
    \caption{The brane perspective uses a different holonomy constraint for RT surfaces than the boundary perspective. \textbf{(a)} RT surfaces can only end on the brane if we compute the entropy for regions $A \cup B$ which include a subregion $B$ of the gravitational part of the brane perspective. This is the usual holonomy constraint at the boundary of AdS. \textbf{(b)} Unlike in the boundary perspective, the RT surface associated to only a subregion $A$ of the asymptotic boundary cannot end on the brane.}
    \label{fig:RT_surface_allowed_disallowed}
\end{figure}
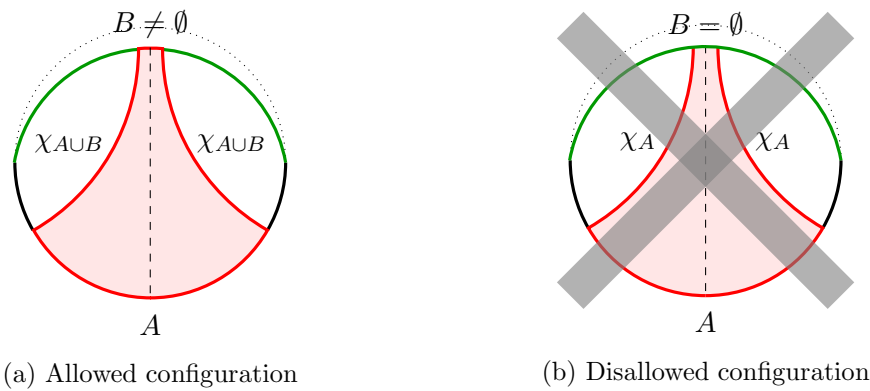

However, an important point is that the homology constraint we need to use to compute entropies in the brane perspective, and consequently the candidate RT surfaces, is different than in the boundary perspective. Instead of \cref{eq:homology_bcft}, the correct homology constraint now is
\begin{align}
    \label{eq:homology_boundary}
        \partial \Sigma_A = A \cup \chi_A.
\end{align}
We can argue in favour of this prescription under some reasonable assumptions. From here on and for the remainder of this paper we will assume the following.
\begin{enumerate}
    \item The theory in the brane perspective is a quantum mechanical theory. Furthermore, there exists a unitary map defined on an appropriate low-energy subset of the Hilbert space which maps a state in the boundary perspective to a state in the brane perspective.
    \item The standard rules of entanglement wedge reconstruction still hold in the brane perspective. In particular we assume that RT surfaces bound the entanglement wedge. Moreover, entanglement wedge complementarity and entanglement wedge nesting hold\footnote{See \cref{sec:mapping} for a discussion of entanglement wedge reconstruction. The entanglement wedge $W_E(A)$ of a boundary subregion $A$ is defined as the domain of dependence $\Sigma_A$ as in \cref{eq:homology_bcft} or \cref{eq:homology_boundary}. Entanglement wedge complementarity means that for pure states $\overline{W_E(A)} = W_E(\bar A)$. Entanglement wedge nesting is the statement that for any two boundary regions $A \subset A'$ we have $W_E(A) \subset W_E(A')$.}.
\end{enumerate}
\begin{figure}[t]
    \centering
    \begin{subfigure}{.45\linewidth}
    \centering
    \begin{tikzpicture}[scale=0.9]
    \draw[black, very thick] (0,0) arc (0:-30:2cm);
    \draw[black, very thick] (-4,0) arc (-180:-150:2cm);
    \draw[dotted] (0,0) arc (0:180:2cm); 
    \fill[white!90!red, very thick] (-2,-2) arc (-90:-30:2cm) arc (-120:-180:3.1cm) arc (85:95:2.035) arc (0:-60:3.1cm) arc (-150:-90:2cm);
    \draw[red, very thick] (-2,-2) arc (-90:-30:2cm) arc (-120:-180:3.1cm);
    \draw[red, very thick] (-2,-2) arc (-90:-150:2cm) arc (-60:0:3.1cm);
    \draw[dashed] (-2,1.7) -- (-2, -2);
    \draw[black!40!green, very thick] (0,0) arc (10:170:2.035cm);
    \node[below] at (-2, -2.1) {$A$};
    \node[above] at (-3, 0) {$\chi_A$};
    \node[above] at (-1, 0) {$\chi_A$};
    \node[above] at (-2, 1.7) {$B \subset \bar A$};
    \fill[white!30!blue, very thick, opacity=.6] (-2,1.7) arc (90:70:1.7cm) arc (0:-180:0.6cm) arc (110:90:1.7cm);
    \draw[black!40!blue, very thick] (-2,1.7) arc (90:70:1.7cm) arc (0:-180:0.6cm) arc (110:90:1.7cm);
    \end{tikzpicture}
    \caption{Entanglement wedge $W_E(A)$}
    \label{fig:proof_a}
    \end{subfigure}    
    \begin{subfigure}{.45\linewidth}
    \centering
    \begin{tikzpicture}[scale=0.9]
    \draw[black, very thick] (-2,-2) arc (-90:-30:2cm);
    \draw[black, very thick] (-2,-2) arc (-90:-150:2cm);
    \draw[dotted] (0,0) arc (0:180:2cm); 
    \draw[red, very thick] (0,0) arc (10:170:2.035cm);
    \fill[white!90!red, very thick, draw = red] (0,0) arc (0:-30:2cm) arc (-120:-180:3.1cm) arc (85:10:2.035cm);
    \fill[white!90!red, very thick, draw = red] (-4,0) arc (180:210:2cm) arc (-60:0:3.1cm) arc (95:170:2.035cm);
    \draw[dashed] (-2,1.7) -- (-2, -2);
    \node[below] at (-2, -2.1) {$A$};
    \node at (-2, 0) {$\chi_{\bar A}$};
    \node[above] at (-2, 1.7) {$B \subset \bar A$};
    \fill[white!30!blue, very thick, opacity=.6] (-2,1.7) arc (90:70:1.7cm) arc (0:-180:0.6cm) arc (110:90:1.7cm);
    \draw[black!40!blue, very thick] (-2,1.7) arc (90:70:1.7cm) arc (0:-180:0.6cm) arc (110:90:1.7cm);
    \end{tikzpicture}
    \caption{Entanglement wedge $\overline {W_E(A)} = W_E(\bar A)$}
    \label{fig:proof_abar}
    \end{subfigure}
    \caption{\textbf{(a)} In the brane perspective, the red region cannot be the entanglement wedge of $A$ if entanglement wedge complementarity is correct. \textbf{(b)} Although $B$ is a subset of $\bar A$, the region that can be reconstructed from $B$ using HKLL (blue shaded region) is not fully contained in the entanglement wedge $\overline {W_E(A)} = W_E(\bar A)$ of $\bar A$ (red shaded region).}
    \label{fig:proof}
\end{figure}
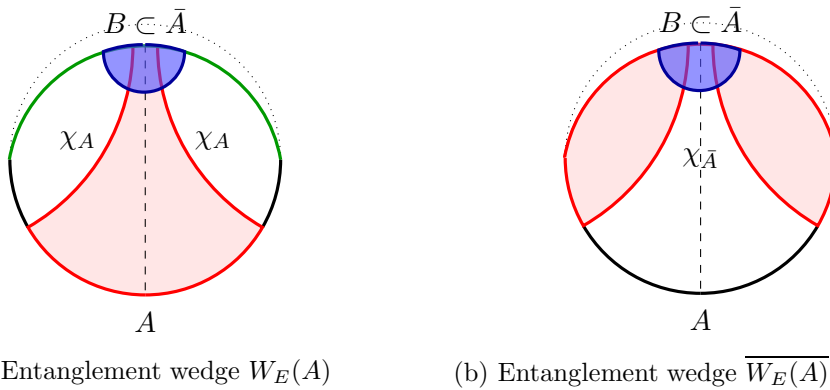
We can argue by contradiction that these assumptions require us to exclude the connected RT surface as a candidate RT surface.
Assume that in the brane perspective the connected surface was the RT surface for a subregion $A$ which is completely contained in the bath, \cref{fig:proof_a}. Since we are in a pure state and have assumed that the boundary theory is a quantum mechanical theory, entanglement wedge complementarity holds. Consider a bulk subregion $\mathcal B$ (e.g., the blue region in \cref{fig:proof}) which intersects $W_E(A)$ and intersects the brane at some $B$ with $\partial \mathcal B \cap \text{brane} = B \subset \bar A$. Any bulk operator in $\mathcal B \cap W_E(A)$ is outside $W_E(\bar A)$ by entanglement wedge complementarity. By entanglement wedge nesting this means that we cannot reconstruct it using boundary operators in $B \subset \bar A$.
This, however, must be wrong. It is known that a slightly modified version of holographic renormalization can be used to build a dictionary for the bulk theory \cite{Gubser:1999vj, PerezVictoria:2001pa, Porrati:2001gx,Neuenfeld:2021wbl}. It was shown in \cite{Neuenfeld:2021wbl} that in the brane perspective an extrapolate dictionary exists which relates bulk fields to boundary operators as well as dynamical brane sources. This statement is time-independent and thus also holds past the Page time. We can thus use HKLL for bulk reconstruction \cite{Hamilton:2006az}, i.e., to reconstruct any operator near the brane from boundary conditions at the brane together with the bulk equations of motion. It follows that the region around the brane must always be in the entanglement wedge of $\bar A$. This is in conflict with entanglement wedge complementarity. We have thus shown by contradiction, that \cref{fig:connected_RT_surface} cannot be a RT surface for a bath subregion $A$, \cref{fig:RT_surface_allowed_disallowed}. 

We can prevent such surfaces from being considered in the RT formula if we use the homology condition that is used for computing RT surfaces in global AdS, \cref{eq:homology_boundary}. In that case, the configuration \cref{fig:RT_surface_disallowed} is not allowed anymore. In practice, this means that in the brane perspective we require a stricter homology condition such that the RT surface must only end on the boundary of the boundary subregion which it is associated to. Let us stress that this in particular means that the RT surface for the brane perspective does not undergo a phase transition at the Page time, unlike the RT surface for the boundary perspective. This also gives rise to an interesting observation: In the brane perspective the gravitating subregion has the associated RT surface shown in \cref{fig:RT_surface_B_big}. The length of this RT surface grows linearly without bound. This is precisely the behaviour of the subregion entropy in the semi-classical description. Lastly, we should point out that this homology constraint is compatible with the results of \cite{Bousso:2001cf} which have argued that the ``holographic domain'' of the brane in the Karch-Randall scenario \cite{Karch:2000ct} should be the red region in \cref{fig:RT_surface_B_big}.

\section{Mapping between the Brane and Boundary Perspective}
\label{sec:mapping}
The homology constraint in the boundary perspective is well motivated in order to correctly reproduce boundary von Neumann entropies. Above we have argued that a different homology constraint should be applied when asking for von Neumann entropies of bath subregions in the brane perspective. 

This leads to a seemingly paradoxical situation when computing the von Neumann entropy of a subregion $A$ of the bath region. Before the Page time the RT surface in both prescriptions is given by the disconnected surface, \cref{fig:disconnected_RT_surface}. After the Page time the entropy in the boundary perspective is given by the connected surface \cref{fig:connected_RT_surface}. This configuration, however, is not allowed in the brane perspective and thus there the correct RT surface is still given by \cref{fig:disconnected_RT_surface}. We must conclude that the entropy of the region $A$ in the boundary perspective is different than the entropy of the region $A$ in the brane perspective.

Our general philosophy will be to assume that the interpretation of the RT formula is unchanged. That is, it computes the von Neumann entropy of quantum fields in the region $A$. We get different answers, however, since the density matrix is encoded in different degrees of freedom in the brane and boundary perspective and we only compute the entropy of the degrees of freedom located in region $A$. This raises the question of how we can translate operators and other quantities between the perspectives. Answering this question is the goal of this section. Before we turn to this, we will review the relationship between extremal surfaces and bulk reconstruction and establish some notation.

\subsection{RT Surfaces and Bulk Reconstruction}
Entanglement wedge reconstruction \cite{Almheiri:2014lwa} is a statement about the encoding of bulk degrees of freedom in the boundary CFT. It is succinctly summarized in the bulk reconstruction theorem of \cite{Dong:2016eik} which states that information about a bulk region $\mathcal A$ can be reconstructed from information in a boundary region $A$, if $\mathcal A$ is contained in the entanglement wedge $W_E(A)$ of $A$. The entanglement wedge $W_E(A)$ is the domain of dependence of a partial Cauchy slice $\Sigma_A$ that connects the boundary region $A$ to the RT surface $\chi_A$ associated with $A$. Here, ``connects'' means that
\begin{align}
    \partial \Sigma_A = A \cup \chi_A \cup X, && X \subset \begin{cases} \text{ETW brane} & \text{(boundary perspective),} \\ \emptyset &\text{(brane perspective).}\end{cases}
\end{align}
Thus, in order to understand which part of the bulk we can access from a subregion $A$ of the boundary we need to understand the different possible bulk RT surfaces $\chi_A$ subject to the homology constraint implemented by $X$ in the above formula. We see that the choice of homology constraint does not only affect the result for holographic entropy computations, but also how the bulk can be reconstructed.

In \cref{fig:RT_surfaces_boundary_perspective,fig:RT_gravity_subregion,fig:RT_surface_allowed_disallowed} the region contained in the respective entanglement wedge was shaded in red. This bulk region is encoded in the quantum state in the associated boundary subregion. Consider a symmetric region $A$ in the bath. After the Page time, boundary and brane perspective yield different entanglement wedges $W_E(A)$. As reviewed above, at sufficiently late times in the boundary perspective, the RT surface connects to the brane such that some information about bulk fields near the brane is in the entanglement wedge. By entanglement wedge reconstruction this information can thus be recovered in the bath region. 
However, using the homology constraint appropriate for the brane perspective the transition of the RT surface never happens. RT surfaces associated with subregions of the bath cannot connect to the brane anymore and the information close to the brane is never accessible in the bath. Let us reiterate this point. In the boundary perspective information about the inside of the black hole is contained in the bath after the Page time, while in the brane perspective it is contained in the gravitational region and \emph{never} accessible from the bath. We must thus conclude that information is encoded differently into the two perspectives. Moreover, the same information appears exactly once in either perspective.

Using entanglement wedge reconstruction one can express a local bulk operator that acts within the entanglement wedge as an operator that acts on the associated boundary subregion. We will mostly not be interested in the details of such reconstruction which can be found in \cite{Dong:2016eik, Faulkner:2017vdd, Cotler:2017erl, Chen:2019gbt}, with one notable exception: Roughly speaking it is much easier to reconstruct operators which are close to the boundary than operators which are deep in the entanglement wedge. We will now make this statement more precise.

\subsection{Extremal Surfaces, Simple and Complex Quantities}
In spacetimes with more than one candidate RT surface the entanglement wedge can be split into two parts. We can consider the \emph{outer wedge} $W_O(A)$ \cite{Engelhardt:2017aux, Engelhardt:2018kcs}, which is the domain of dependence of a partial Cauchy slice which connects the outermost RT candidate surface with the boundary region $A$. In our case and from the point of view of the boundary perspective, we find that before the Page time, the outer and entanglement wedges of a bath subregion $A$ agree, \cref{fig:disconnected_RT_surface}. However, after the Page time, the outer wedge is a proper subset of the entanglement wedge, \cref{fig:connected_RT_surface}. The region in the bulk that is not part of the outer wedge is called the \emph{(Python's) lunch} \cite{Brown:2019rox}.

It is relatively easy to reconstruct operators in the outer wedge, as was shown in \cite{Engelhardt:2021mue}. There, the authors define the \emph{simple wedge} $W_S(A)$ as the bulk region that can be reconstructed from one-point functions of simple operators with arbitrary simple sources turned on, together with time evolution on multiple time folds. 
Simple sources are by definition sources that create excitations which causally propagate into the bulk, while simple operators are those which couple to infinitesimal simple sources. For example, sources that couple to relevant single-trace operators are simple. However, sources which couple to operators that acausally create excitations deep in the bulk are generally not simple. The main result of \cite{Engelhardt:2021mue} is that the simple wedge and the outer wedge agree (see also \cite{Levine:2020upy}).

In our case, this is very easy to see. Since we are in the global AdS vacuum, the outer wedge agrees with the causal wedge. This implies that, at least in the limit of vanishing backreaction, fields in the outer wedge can be reconstructed by methods introduced by HKLL \cite{Hamilton:2006az}. One expands bulk fields in the causal wedge and uses the extrapolate dictionary to relate them to CFT operators. By inverting this relation bulk fields can be written as
\begin{align}
    \phi(x) = \int_A K(x,y) \mathcal O(y).
\end{align}
Thus, simple operators are those which only depend on bulk data in the outer wedge. In the following, by \emph{simple quantities of} $A$ we will mean all quantities derived from simple operators and operator expectation values within $A$.

We define \emph{complex operators of} $A$ as those whose bulk representation is not fully contained in $W_O(A)$. This definition agrees with the point of view taken in \cite{Brown:2019rox}, where it was argued that in order to reconstruct bulk information contained in the Python's lunch, one needs a quantum circuit with exponentially many gates. For us, the most important example of a complex quantity is the reduced density matrix $\rho_A$ of a bath subregion $A$ in the boundary perspective after the Page time, where the entanglement wedge extends beyond the outer wedge. Of course, all derived quantities which depend on fine-grained knowledge of $\rho_A$ are also complex. The prime example is $S_\text{vN}(A)$ in the boundary perspective after the Page time. Another interesting class of complex quantities are operators which probe the inside region of an evaporating black hole after the Page time by acting on the radiation\footnote{In fact, in the case of an eternal black hole this also includes operators just outside the black hole horizon.}. As we will see, such operators can be reconstructed in the boundary perspective as operators in a sufficiently large bath subregion\footnote{In \cite{Kim:2020cds} similar ideas were explored from an information theoretic point of view.}. Again the term \emph{complex quantities} includes all derived quantities, such as an observer which controls the Hawking radiation past the Page time to a degree sufficient to extract information about the black hole interior.

Before we go on, we should caution the reader that the results of \cite{Brown:2019rox, Engelhardt:2021mue} were obtained for the classical case and compact entangling surfaces. Here, we simply assume without proof that the conclusions essentially carry over if we consider entangling surfaces anchored at the boundary as well as quantum corrections \cite{Engelhardt:toappear}. And in fact, as discussed in \cite{Engelhardt:2021mue, Engelhardt:private} this is expected.

\subsection{Mapping Operators between the two Perspectives}
We will now turn to the question of how operators map between the brane and the boundary perspective. The upshot of the following will be that simple bath operators are identical in the boundary and brane perspective. Once we start asking fine-grained questions the situation is more complicated. We will see that complex quantities translate non-trivially between the boundary and brane perspective.

\begin{figure}[t]
    \centering
    \begin{subfigure}{.31\linewidth}
    \centering
    \begin{tikzpicture}[scale=0.9]
    \draw[dotted] (0,0) arc (0:180:2cm); 
    \draw[very thin, gray] (0,0) arc (10:170:2.035cm);
    \draw[black, very thick] (0,0) arc (0:-30:2cm);
    \draw[black, very thick] (-4,0) arc (-180:-150:2cm);
    \fill[black!40!green, draw=black] (-0.1,-0.1) rectangle (0.1,0.1);
    \fill[black!40!green, draw=black] (-4.1,-0.1) rectangle (-3.9,0.1);
    \fill[white!90!red, draw=red, very thick] (-2,-2) arc (-90:-30:2cm) arc (55:125:3.03cm) arc (-150:-90:2cm);
    \draw[dashed] (-2,1.7) -- (-2, -2);
    \node[below] at (-2, 0) {$\chi_A$};
    \fill[draw=black] (-1.9,-1.9) circle (0.1cm) -- (-1,-2.2) node [right] {$\phi_S(x)$};
    \node[below] at (-2, -2.1) {$A$};
    \end{tikzpicture}
    \caption{Boundary perspective, $t < t_P$}
    \label{fig:a_simple}
    \end{subfigure}
    \begin{subfigure}{.31\linewidth}
    \centering
    \begin{tikzpicture}[scale=0.9]
    \draw[black, very thick] (0,0) arc (0:-30:2cm);
    \draw[black, very thick] (-4,0) arc (-180:-150:2cm);
    \draw[dotted] (0,0) arc (0:180:2cm); 
    \draw[very thin, gray] (0,0) arc (10:170:2.035cm);
    \fill[black!40!green, draw=black] (-0.1,-0.1) rectangle (0.1,0.1);
    \fill[black!40!green, draw=black] (-4.1,-0.1) rectangle (-3.9,0.1);
    \fill[white!98!red, very thick] (-2,-2) arc (-90:-30:2cm) arc (-120:-180:3.1cm) arc (85:95:2.035) arc (0:-60:3.1cm) arc (-150:-90:2cm);
    \fill[white!90!red, dotted, draw=red] (-2,-2) arc (-90:-30:2cm) arc (55:125:3.03cm) arc (-150:-90:2cm);
    \draw[red, very thick] (-2,-2) arc (-90:-30:2cm) arc (-120:-180:3.1cm);
    \draw[red, very thick] (-2,-2) arc (-90:-150:2cm) arc (-60:0:3.1cm);
    \draw[dashed] (-2,1.7) -- (-2, -2);
    \node[below] at (-2, -2.1) {$A$};
    \node[above] at (-3, 0) {$\chi_A$};
    \node[above] at (-1, 0) {$\chi_A$};
    \fill[draw=black] (-1.9,-1.9) circle (0.1cm) -- (-1,-2.2) node [right] {$\phi_S(x)$};
    \end{tikzpicture}
    \caption{Boundary perspective, $t > t_P$}
    \label{fig:b_simple}
    \end{subfigure}
    \begin{subfigure}{.31\linewidth}
    \centering
    \begin{tikzpicture}[scale=0.9]
    \draw[dotted] (0,0) arc (0:180:2cm);
    \draw[black, very thick] (0,0) arc (0:-30:2cm);
    \draw[black, very thick] (-4,0) arc (-180:-150:2cm);
    \fill[black!40!green, draw=black] (-0.1,-0.1) rectangle (0.1,0.1);
    \fill[black!40!green, draw=black] (-4.1,-0.1) rectangle (-3.9,0.1);
    \fill[white!90!red, draw=red, very thick] (-2,-2) arc (-90:-30:2cm) arc (55:125:3.03cm) arc (-150:-90:2cm);
    \draw[black!40!green, very thick] (0,0) arc (10:170:2.035cm);
    \draw[dashed] (-2,1.7) -- (-2, -2);
    \node[below] at (-2, -2.1) {$A$};
    \node[below] at (-2, 0) {$\chi_A$};
    \fill[draw=black] (-1.9,-1.9) circle (0.1cm) -- (-1,-2.2) node [right] {$\phi_S(x)$};
    \end{tikzpicture}
    \caption{Brane perspective, any $t$}
    \label{fig:c_simple}
    \end{subfigure}
    \caption{A field $\phi_S(x)$ dual to a simple operator $\mathcal O_S$ in a bath subregion $A$ is encoded in the boundary and brane perspective in the same way. In all three cases \textbf{(a)},\textbf{(b)},\textbf{(c)} the operator lies in the outer wedge $W_O(A)$ (region shaded in darker red) associated to $A$ and the operator can be reconstructed using HKLL. This is only possible because the operator is not in the lunch (lighter shaded region in \textbf{(b)}).}
    \label{fig:operator_reconstruction_simple}
\end{figure}
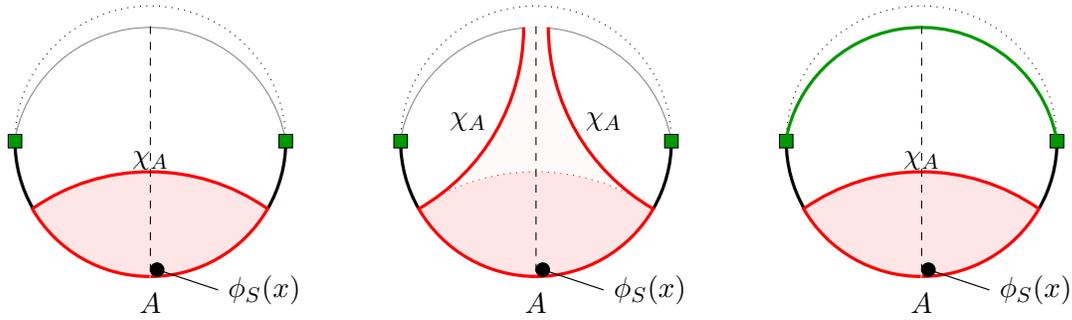

Let us first consider a simple operator $\mathcal O_S$ in a bath subregion $A$, \cref{fig:operator_reconstruction_simple}. By definition, simple means that from the bulk point of view, the operator can be written in the bulk given access to only the the outer wedge $W_O(A)$. Before the Page time, the entanglement wedge $W_E(A)$ in the brane and the boundary perspectives agrees with the outer wedge, and the outer wedges agree between the two perspectives. After the Page time, $W^\text{boundary}_E(A) \neq W^\text{brane}_E(A)$. However, for the outer wedges we still have that $W^\text{boundary}_O(A) = W^\text{brane}_O(A)$. Since the HKLL procedure to reconstruct the simple operator in the outer wedge is identical in both perspectives we find that simple operators in the bath map trivially between the brane and the boundary perspective.

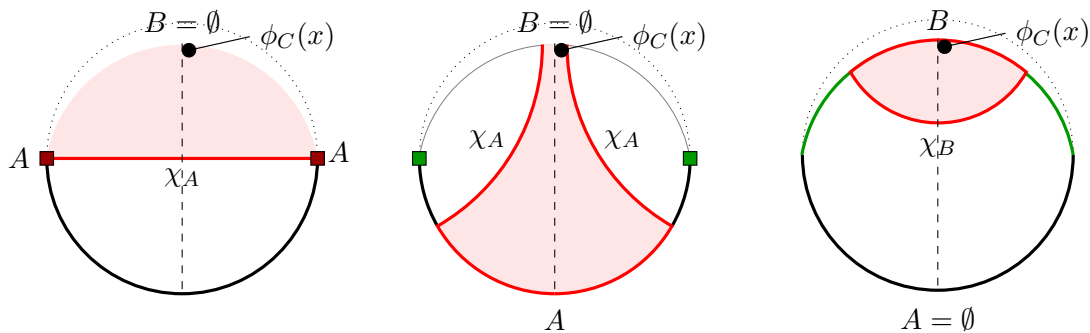
\begin{figure}[t]
    \centering
    \begin{subfigure}{.31\linewidth}
    \centering
    \begin{tikzpicture}[scale=0.9]
    \draw[black, very thick] (0,0) arc (0:-180:2cm);
    \draw[dotted] (0,0) arc (0:180:2cm); 
    \fill[very thin, white!90!red] (0,0) arc (10:170:2.035cm) -- (0,0);
    \draw[red, very thick] (-4,0) -- (0,0);
    \fill[black!40!red, draw=black] (-0.1,-0.1) rectangle (0.1,0.1);
    \fill[black!40!red, draw=black] (-4.1,-0.1) rectangle (-3.9,0.1);
    \draw[dashed] (-2,1.7) -- (-2, -2);
    \node[left] at (-4.1, 0) {$A$};
    \node[right] at (0, 0.1) {$A$};
    \node[above] at (-2, 1.7) {$B = \emptyset$};
    \node[below] at (-2, 0) {$\chi_A$};
    \fill[draw=black] (-1.9,1.6) circle (0.1cm) -- (-1,1.8) node [right] {$\phi_C(x)$};
    \node[below, opacity=0] at (-2, -2.1) {$A$};
    \end{tikzpicture}
    \caption{Boundary perspective, $t < t_P$}
    \label{fig:b1}
    \end{subfigure}
    \begin{subfigure}{.31\linewidth}
    \centering
    \begin{tikzpicture}[scale=0.9]
    \draw[black, very thick] (0,0) arc (0:-30:2cm);
    \draw[black, very thick] (-4,0) arc (-180:-150:2cm);
    \draw[dotted] (0,0) arc (0:180:2cm); 
    \draw[very thin, gray] (0,0) arc (10:170:2.035cm);
    \fill[black!40!green, draw=black] (-0.1,-0.1) rectangle (0.1,0.1);
    \fill[black!40!green, draw=black] (-4.1,-0.1) rectangle (-3.9,0.1);
    \fill[white!90!red, very thick] (-2,-2) arc (-90:-30:2cm) arc (-120:-180:3.1cm) arc (85:95:2.035) arc (0:-60:3.1cm) arc (-150:-90:2cm);
    \draw[red, very thick] (-2,-2) arc (-90:-30:2cm) arc (-120:-180:3.1cm);
    \draw[red, very thick] (-2,-2) arc (-90:-150:2cm) arc (-60:0:3.1cm);
    \draw[dashed] (-2,1.7) -- (-2, -2);
    \node[below] at (-2, -2.1) {$A$};
    \node[above] at (-2, 1.7) {$B = \emptyset$};
    \node[above] at (-3, 0) {$\chi_A$};
    \node[above] at (-1, 0) {$\chi_A$};
    \fill[draw=black] (-1.9,1.6) circle (0.1cm) -- (-1,1.8) node [right] {$\phi_C(x)$};
    \end{tikzpicture}
    \caption{Boundary perspective, $t > t_P$}
    \label{fig:b2}
    \end{subfigure}
    \begin{subfigure}{.31\linewidth}
    \centering
    \begin{tikzpicture}[scale=0.9]
    \draw[black, very thick] (0,0) arc (0:-180:2cm);
    \draw[dotted] (0,0) arc (0:180:2cm); 
    \draw[black!40!green, very thick] (0,0) arc (10:170:2.035cm);
    \fill[white!90!red, draw=red, very thick] (-2,1.7) arc (90:50:2.035cm) arc (-30:-150:1.5cm) arc (130:90:2.035cm);
    \draw[dashed] (-2,1.7) -- (-2, -2);
    \node[below] at (-2, -2.1) {$A = \emptyset$};
    \node[above] at (-2, 1.7) {$B$};
    \node[below] at (-2, 0.4) {$\chi_B$};
    \fill[draw=black] (-1.9,1.6) circle (0.1cm) -- (-1,1.8) node [right] {$\phi_C(x)$};
    \end{tikzpicture}
    \caption{Brane perspective, any $t$}
    \label{fig:a}
    \end{subfigure}
    \caption{A bulk operator $\phi_C(x)$ is encoded differently in the boundary and brane perspective. \textbf{(a)} In the boundary perspective before the Page time the operator $\phi_C(x)$ can be expressed in terms of boundary degrees of freedom. \textbf{(b)} In the boundary perspective after the Page time, a large subregion $A$ of the bath must be used to reconstruct $\phi_C(x)$, since it is not accessible from boundary degrees of freedom anymore. It is now dual to a complex operator $\mathcal O_C$. \textbf{(c)} In the brane perspective, a subregion $B$ of the gravitational region is sufficient to reconstruct $\phi_C(x)$. If we had only access to the bath, we could not reconstruct $\phi_C(x)$, c.f., \cref{fig:RT_surface_allowed_disallowed}. This is true before and after the Page time.}
    \label{fig:operator_reconstruction}
\end{figure}

Now, let us consider the case of a a complex operator $\mathcal O_C$ in the boundary perspective, \cref{fig:operator_reconstruction}. In the boundary perspective its representation is time-dependent. Before the Page time, we can reconstruct the operator in terms of defect operators in the BCFT. After the Page time, we can write it in terms of fields in region $A$. Since it is not located in the outer wedge $W_O(A)$, but in the lunch, this operator is complex. In the brane perspective the operator $\mathcal O_C$ can \emph{only} be represented as an operator in the region with dynamical gravity. However, here it is a simple operator---independent of time. Furthermore, note that as we move the operator towards the brane, it becomes a local operator in the gravitational region of the brane perspective \cite{Neuenfeld:2021wbl}. We see that in the brane perspective it can be represented as a local field on the brane. 

More generally, in the brane perspective, any bulk operator contained in the red region in \cref{fig:RT_surface_B_big} is a simple operator for some subregion $B$ on the brane. Any operator contained in the complement of the red region in \cref{fig:RT_surface_B_big} can be written as a simple operator for some bath subregion $A$. In contrast, in the boundary perspective, we can have operators in the red region of \cref{fig:RT_surface_B_big} that can be reconstructed using a proper\footnote{That is, the bath subregion does not include the defects at the BCFT boundary.} bath subregion $A$. Such operators will always be complex, if we only have access to $A$. All bulk operators contained in the complement of the red region in \cref{fig:RT_surface_B_big} will be simple for some $A$ which is a proper bath subregion.

This demonstrates that there is a non-trivial dictionary between the brane and the bulk perspective which moreover is time-dependent. Furthermore, this nicely illustrates one of our main results, namely that information is encoded only once, but might be encoded in different degrees of freedom in the brane and the boundary perspective. The benefit of the doubly-holographic scenario at hand is that at least in principle, this dictionary can be found explicitly. Given any bulk excitation, we can relate it to an excitation in the BCFT as well as the boundary perspective via the appropriate holographic dictionaries.

\subsection{Mapping Entropies between the two Perspectives}
As we have seen above, after the Page time the von Neumann entropy of a bath subregion $A$ differs between the two perspectives. We will now investigate how to compute the von Neumann entropy in the brane perspective from the boundary perspective and vice versa. 

Let us first focus on how to obtain the boundary perspective answer (the connected surface) from the brane perspective. The answer to this question was given in \cite{Chen:2020uac}. If we consider all additional regions $B$ in the gravitational region of the brane perspective and compute
\begin{align}
\label{eq:island}
S_\text{island}(A) = \underset{B}{\min \text{ext}} \left\{ \frac{\text{Area}(\partial B)}{4 \Gbrane} + S_\text{vN}(A \cup B)\right\},
\end{align}
we exactly reproduce the entropy computed in the boundary perspective. If we compute \cref{eq:island} using the RT prescription for the brane perspective, the resulting RT surface is exactly the one that computes $S_\text{vN}(A)$ in the boundary perspective. We conclude that
\begin{align}
S_\text{island}(A) \text{ in the \textbf{brane} perspective} = S_\text{vN}(A) \text{ in the \textbf{boundary} perspective.}
\end{align}
\Cref{eq:island} is of course the island formula which was given in \cite{Almheiri:2019hni} and conjectured to compute the fine-grained entropy of Hawking radiation. In the doubly-holographic setting this conjecture is correct if we identify the fine-grained entropy with the von Neumann entropy in the boundary perspective.

In order to compute the von Neumann entropy of the same subregion in the brane perspective from the boundary perspective, remember that in the boundary perspective the disconnected surface is still an extremal surface, although not the minimal one. Instead, it is the outermost extremal surface, which was used in the definition of the outer wedge $W_O(A)$. In \cite{Engelhardt:2021mue}, based on \cite{Engelhardt:2017aux, Engelhardt:2018kcs}, it was argued that the area of the outermost extremal surface is a coarse-grained entropy $S_\text{outer}$ which is a measure for the number of states whose outer wedges agree with the outer wedge of $A$, while the lunches might be arbitrary (but still obey Einstein's equations).

It was proposed in \cite{Engelhardt:2017aux, Engelhardt:2018kcs} and proven in \cite{Engelhardt:2021mue} that the outer entropy is holographically dual to \emph{simple entropy} in the CFT. Simple entropy is defined as a measure for the number of states that reproduce one-point functions of simple operators in the presence of simple sources
\begin{align}
    \label{eq:simple_entropy_region}
    S_\text{simple}(A) = \max_{\rho_A'} \left(S_\text{vN}[\rho_A']: \langle E^\dagger \mathcal O(t) E \rangle_{\rho_A'} = \langle E^\dagger \mathcal O(t) E \rangle_{\rho_A} \right), && E = \mathcal T e^{i \int_A dx J(x)\mathcal O(x)}.
\end{align}
In general, as argued in \cite{Engelhardt:2021mue}, the domain of integration in the definition of $E$ should involve many time-folds. However, in our doubly-holographic setup this is not necessary.
It is clear from this definition that the simple entropy is again a course-grained entropy. \Cref{eq:simple_entropy_region} measures the number of states which reproduce prescribed correlation functions for all HKLL-type boundary operators sandwiched between forward and backwards time evolution with simple sources turned on. Correlation function of more complicated operators (i.e., operators living in the Python's lunch) need not agree for those states and can only be reproduced from the true state $\rho_A$. Extrapolating the results of \cite{Engelhardt:2017aux, Engelhardt:2018kcs, Engelhardt:2021mue} to extremal surfaces which can be anchored on the boundary we thus have that for a bath subregion $A$
\begin{align}
S_\text{vN}(A) \text{ in the \textbf{brane} perspective} = S_\text{simple}(A) \text{ in the \textbf{boundary} perspective,}
\end{align}
i.e., the brane perspective von Neumann entropy is a coarse grained entropy in the boundary perspective.

Next, we will show using a simple toy model, how a mapping between two perspectives naturally gives rise to an island formula and wormholes. Before we do this, however, let us comment on the island formula. The microscopic origin of the island formula is not understood in dimensions greater than two. In two dimensions, it is explained through ensemble averaging of theories. However, this explanation does not seem to extend easily to higher dimensions, since it is generally believed that canonical example of holography such as $\mathcal N=4$ SYM theory are essentially unique. Our result points in a different direction. Our claim is that the island formula can be used to compute the entropy of the density matrix of a subregion in a quantum gravitational formulation of a theory from a semi-classical computation. However, it is purely a statement about entropies in the quantum gravitational formulation and not about entropies in the semi-classical approximation. In particular---we will further comment on this below---in our case the quantum gravitational formulation consists of a bath coupled to a defect and does not contain an easily accessible description (i.e., a geometric description) of the black hole interior. Rather, the interior is encoded in complex correlations.

\section{A Qubit Model for Islands and Wormholes}
\label{sec:model}
We have seen that in doubly-holographic models we can compute the von Neumann entropy in the boundary perspective by using the island entropy functional in the brane perspective. The two perspectives are related by reorganizing information about the state into different degrees of freedom. In this section we will show, using a simple qubit toy model, that if the reorganization of degrees of freedom is sufficiently complicated, an island formula emerges. Moreover, we will also see behaviour that can be interpreted as the appearance of wormholes.

\subsection{Von Neumann Entropy and Islands}
Consider a pure state black hole with a coarse grained entropy of $n$ bits which after evaporation has yielded $n$ qubits of Hawking radiation. We denote the Hilbert space of the radiation as $\mathcal H_R$ and say that the number of qubits in $R$ is $|R|=n$. Since the black hole dynamics has scrambled the information of the state that formed the black hole, all subsystems $A$ of the radiation with $m < \frac n 2$, where $m = |A|$, are almost maximally entangled with the rest of the radiation contained in $\bar A$, with $|\bar A| = n-m$. The reduced density matrix on $A$ is given by
\begin{align}
    \rho_A = \tr_{\bar A} \rho
\end{align}
and its entropy behaves as
\begin{align}
    \label{eq:entropy_radiation}
    S_\text{vN}(\rho_A) = \begin{cases} m \text{ bit} & \text{if } m < \frac {n} 2,\\ m-n \text{ bit} & \text{if }  \frac {n} 2 \leq m \leq n.
    \end{cases}
\end{align}
This is the von Neumann entropy of a subset of the Hawking radiation. We call the $m$-dependence of the von Neumann entropy the Page curve of our model. We might think of $\rho$ as the state in the true quantum gravitational description of our system, i.e., the boundary perspective. In particular, a small subsystem has a maximally mixed density matrix and looks like a thermal state. In our case the state has been prepared by some random unitary such that a small subsystem is at infinite temperature.

\begin{figure}
    \centering
    \begin{tikzpicture}
        \draw (-5,0) rectangle (-4.5,.5) rectangle (-4,0) rectangle (-3.5,.5);
        \draw (-2.5,0) rectangle (-2,.5);
        \draw[decorate, decoration={brace, amplitude=10pt}] (-5,.6) -- node[above, yshift=10pt]{$\mathcal H_R$, $n$ qubits} (-2,.6);
    
        \draw[->] (-1.5, 0.25) -- (-0.5,0.25);
        \draw (0,0) rectangle (.5,.5) rectangle (1,0) rectangle (1.5,.5);
        \draw (2.5,0) rectangle (3,.5);
        \draw[decorate, decoration={brace, amplitude=10pt}] (0,.6) -- node[above, yshift=10pt]{$\mathcal H_R$, $n$ qubits} (3,.6);
        
        \draw (3.25,0) rectangle node {0} (3.75,.5) rectangle node {0} (4.25,0) rectangle node {0} (4.75,.5);
        \draw (5.75,0) rectangle node {0} (6.25,.5);
        \draw[decorate, decoration={brace, amplitude=10pt}] (3.25,.6) -- node[above, yshift=10pt]{$\mathcal H_I$, $k$ qubits} (6.25,.6);
    \end{tikzpicture}
    \caption{In the boundary perspective (left), the information is completely contained in $\mathcal H_R$. To go to the brane prescription, we enlarge the Hilbert space by adding $k$ qubits is some pure state (right) and scramble the system (not shown).}
    \label{fig:hilbert_spaces}
\end{figure}
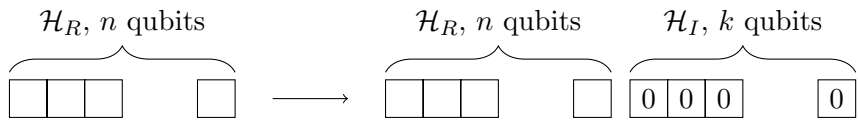

Let us now consider what happens if the state of the radiation is embedded into a bigger Hilbert space $\mathcal H_R \to \mathcal H_R \otimes \mathcal H_I$, \cref{fig:hilbert_spaces}, where $\mathcal H_I$ has dimension $2^k$, i.e. $|I| = k$, and should be thought of as the interior of the black hole. We assume that this encoding happens with a very complex operation which we model by a random unitary, such that
\begin{align}
    \label{eq:scrambled_density_matrix}
    \tilde \rho = U (\rho \otimes | 0 \rangle \langle 0 |^{\otimes k}) U^\dagger.
\end{align}
Morally, $U$ plays the role of the map between the brane and boundary perspective in the preceding sections. We model $U$ by a complex operator, since it relates simple operators in the brane perspective to complex operators in the boundary perspective. We will think of $\tilde \rho$ as describing our state in the semi-classical description, i.e., the brane perspective. The philosophy is that effective field theory can be understood as a complex reorganization of a smaller number of quantum gravitational degrees of freedom. If $k > n$, as we will assume in the following, the von Neumann entropy of the first $m< n$ qubits of the ``boundary perspective'' description is
\begin{align}
    \label{eq:entropy_effective}
    S_\text{vN}(\tilde \rho_A) = m \text{ bit}, \qquad \text{for }  0 \leq m \leq n && \tilde \rho_A = \tr_{\bar A \cup I} \tilde \rho .
\end{align}
In other words, the Hawking radiation for any $m$ qubits looks completely thermal. This is of course different to the von Neumann entropy of the original state, \cref{eq:entropy_radiation}. However, expectation values of simple observables which only depend on a small number of qubits do not depend on whether we take our state to be $\rho_m$ or $\tilde \rho_m$.

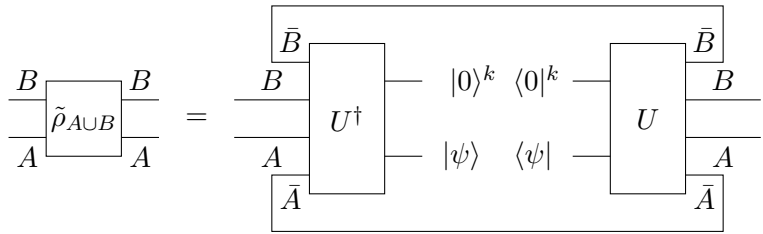
\begin{figure}
    \centering
    \begin{tikzpicture}
        \draw (-4,-1.5) rectangle node {$\tilde \rho_{A \cup B}$} (-3,-0.5);
        \draw (-4.5, -0.75) -- node [above] {$B$} (-4,-0.75);
        \draw (-4.5, -1.25) -- node [below] {$A$} (-4,-1.25);
        \draw (-2.5, -0.75) -- node [above] {$B$} (-3,-0.75);
        \draw (-2.5, -1.25) -- node [below] {$A$} (-3,-1.25);
        \node at (-2,-1) {$=$};
        \draw (-0.5,-2) rectangle node {$U^\dagger$} (0.5,0);
        
        \draw (-1.5, -0.75) -- node [above] {$B$} (-0.5,-0.75);
        \draw (-1.5, -1.25) -- node [below] {$A$} (-0.5,-1.25);
        \draw (0.5, -0.5) -- (1, -0.5);
        \draw (0.5, -1.5) -- (1, -1.5);
        \node at (2.1,-0.5) {$|0\rangle^k \;\; \langle0|^k$};
        \node at (2,-1.5) {$|\psi\rangle \;\;\;\; \langle\psi|$};
        \draw (3.5, -0.5) -- (3, -0.5);
        \draw (3.5, -1.5) -- (3, -1.5);
        \draw (3.5,-2) rectangle node {$U$} (4.5,0);
        \draw (-0.5,-0.25) -- node [above] {$\bar B$} (-1, -0.25) -- (-1, 0.5) -- (5, 0.5) -- (5,-0.25) -- node [above] {$\bar B$} (4.5,-0.25);
        
        \draw (4.5, -0.75) -- node [above] {$B$} (5.5,-0.75);
        \draw (4.5, -1.25) -- node [below] {$A$} (5.5,-1.25);
        \draw (-0.5,-1.75) -- node [below] {$\bar A$} (-1, -1.75) -- (-1, -2.5) -- (5, -2.5) -- (5,-1.75) -- node [below] {$\bar A$} (4.5,-1.75);
        
    \end{tikzpicture}
    \caption{The reduced density matrix used in the toy model island formula. We trace over $\bar A$, that is $n-m$ of the original radiation qubits, as well as over $\bar B$, that is $k-i$ of the added qubits. The $i$ qubit subsystem $B$ can be interpreted as an island or the interior of the evaporated black hole.}
    \label{fig:reduced_density_matrix}
\end{figure}

As we argued before, the island rule tries to reproduce \cref{eq:entropy_radiation} from the density matrix \cref{eq:scrambled_density_matrix}. It can be modeled by replacing $\tilde \rho_A \to \tilde \rho_{A \cup B}$, where $\tilde \rho_{A \cup B}$ is a density matrix on $A$, the first $m \leq n$ qubits, as well as $B$, the last $i \leq k$ qubits, see \cref{fig:reduced_density_matrix}. That is, we have $I = B \cup \bar B$. Replacing $\tilde \rho_A \to \tilde \rho_{A \cup B}$ means we allow to replace our reduced density matrix by a different matrix that apart from the $m = |A|$ qubits in the radiation also describes $i = |B|$ additional qubits in the interior of the black hole. We then \emph{define} a new \emph{qubit island entropy}
\begin{align}
    \label{eq:island_qubit_rule}
    \tilde S_\text{island}(\tilde \rho_A) \equiv \min_{B \subset I} S_\text{vN}(\tilde \rho_{A \cup B}), \qquad \tilde \rho_{A \cup B} = \tr_{\bar A \cup \bar B} \tilde \rho.
\end{align}
The minimization is over choices of the number of qubits in the interior. Of course, this entropy does not really compute the true fine-grained entropy of the first $m$ qubits in the state $\tilde \rho$. This is done by \cref{eq:entropy_effective}. However, it agrees with the von Neumann entropy of the first $m$ qubits in the state $\rho$: A simple computation shows that as long as $m < \frac n 2$ the minimum is taken by considering no additional qubits in the interior. On the other hand, if $m > \frac n 2$ the minimization requires to choose the set $B$ to be all last $k$ qubits, i.e., the full interior of the black hole. The resulting entropy then becomes
\begin{align}
    \tilde S_\text{island}(\tilde \rho_A) = \begin{cases} m \text{ bit} & \text{if } m < \frac n 2,\\ n-m \text{ bit} & \text{if }  \frac {n} 2 \leq m \leq n,
    \end{cases}
\end{align}
which precisely reproduces \cref{eq:entropy_radiation}. We thus see that the qubit island rule \eqref{eq:island_qubit_rule} reproduces the true entropy of $m$ qubits of Hawking radiation using the density matrix \eqref{eq:scrambled_density_matrix}.

\subsection{Purity and Replica Wormholes}
Although the qubit model introduced above is very crude, we can also reproduce a behaviour which is very reminiscent of replica wormholes. In the path integral formulation of semi-classical gravity wormholes appear in replica geometries used to compute entropies. In the limit of a single replica, they give rise to the island formula.

We thus also expect a sign of replica wormholes in the computation of the purity,
\begin{align}
    \mu(\rho_A) = \tr \rho_A^2.
\end{align}
Here, we will keep the notation of the previous subsection. Again, we assume that $\rho$ is the state of Hawking radiation after the black holes has completely evaporated. Information about the initial state of matter which formed the black hole is completely scrambled. As usual, we will model the scrambling by a Haar random unitary.

We can compute the perspective in the ``boundary perspective'' by assuming that the state $\rho$ is a generic pure state
\begin{align}
    \rho = V | 0 \rangle \langle 0| ^{\otimes n}V^\dagger
\end{align}
and using the formula for Haar averaging over unitary operators
\begin{align}
    \label{eq:haar_avg}
    \int dU \; U_{i_1j_1}U_{i_2j_2}U^*_{i'_1j'_1}U_{i'_2j'_2} = \frac1 {d^2 - 1} \left( \delta_{i_1 i'_1} \delta_{i_2 i'_2} \delta_{j_1 j'_1} \delta_{j_2 j'_2} + (i'_1, j'_1 \leftrightarrow i'_2, j'_2)\right) + \mathcal O(d^{-3}).
\end{align}
To leading order the resulting purity is
\begin{align}
    \label{eq:renyi_radiation}
    \mu(\rho_A) \simeq \frac 1 {2^m} + \frac 1 {2^{n-m}} \simeq \begin{cases} \frac 1 {2^m} & \text{if } m < \frac n 2,\\ \frac 1 {2^{n-m}} & \text{if }  \frac n 2 \leq m \leq n.
    \end{cases}.
\end{align}
Note that we do not want to interpret the result as a result obtained by actually averaging over all states $\rho$. Rather, we simply assume that $\rho$ is fixed and behaves like a typical state.

Again, we want to reproduce this result from the effective description introduced above, i.e., we encode the state $\rho$ into a bigger Hilbert space using a random unitary, \cref{eq:scrambled_density_matrix}. A computation of purity of the first $m < n$ qubits in the effective description, now using \cref{eq:haar_avg} for the encoding unitary $U$, yields
\begin{align}
    \mu(\rho_A) = \frac 1 {2^m} + \frac 1 {2^{k + n - m}} \simeq \frac 1 {2^m}.
\end{align}
In order to approximately reproduce \cref{eq:renyi_radiation} we again \emph{define} the \emph{wormhole purity}
\begin{align}
    \label{eq:wormhole_purity}
    \tilde \mu_\text{wormhole}(\tilde \rho_A) = \max_{B \subset I} \mu(\tilde \rho_{A \cup B}).
\end{align}
To motivate the maximization procedure note that the island entropy instructs us to minimize a certain von Neumann entropy, which can be obtained as a limit of Renyi entropies. Purity is related to the second Renyi entropy $S_2$ via $\mu = \exp(-S_2)$. Thus maximizing purity implies minimizing Renyi entropy. Whenever the maximization procedure on the right hand side instructs to use $\tilde \rho_{A \cup B}$ with a non-empty set $B$, we can interpret the computation as forming a wormhole in the black hole interior. The reason is that we can describe the doubled system outside the black hole by the non-factorizable density matrix
\begin{align}
    \tilde \rho_{A,A'} = \tr_{B} (\tilde \rho_{A \cup B} \otimes \tilde \rho_{A' \cup B}),
\end{align}
where $B$ is the set that minimizes \cref{eq:wormhole_purity}.
Using the wormhole purity, we obtain 
\begin{align}
    \tilde \mu_\text{wormhole}(\tilde \rho_A) = \max_{B} \left(\frac 1 {2^{m + i}} + \frac 1 {2^{n-m + k - i}} \right) = \begin{cases} \frac 1 {2^m} & \text{if } m < \frac n 2,\\ \frac 1 {2^{n-m}} & \text{if }  \frac n 2 \leq m \leq n, \end{cases}
\end{align}
where the upper case is given by $i = 0$, and in the lower case is given by $i = k$.
This approximately reproduces \cref{eq:renyi_radiation} as claimed.

\section{Black Hole Complementarity}
\label{sec:bhc}
\subsection{The Postulates of BHC}
Black Hole Complementary (BHC) is a framework which attempts to describe black holes as quantum mechanical systems, while at the same time being compatible with semi-classical gravity. It posits that from afar a black hole obeys certain quantum mechanical, relativistic and statistical conditions \cite{Susskind:1993if}: For an observer far away from the black hole it can be described as a quantum mechanical system with unitary time evolution. Away from the horizon semi-classical gravity is a good approximation to the dynamics. In particular the process of how Hawking radiation gets transported away from the horizon can be described within the framework of semi-classical gravity. Lastly, it is assumed that the black hole entropy is a coarse grained entropy, i.e., there are $\sim \exp(S(M))$ black hole microstates compatible with a classical black hole of mass $M$ and entropy $S(M)$. From the point of view of string theory, and in particular the AdS/CFT correspondence these postulates seem to be realized, and thus do not seem particularly problematic.

And in fact, a problem with these postulates arises only if the semi-classical description of gravity remains valid at and beyond the black hole horizon. For a sufficiently large black hole, there is no indication within semi-classical gravity that it ceases to be valid. Taking the view of an infalling observer illustrates this very clearly: For a sufficiently large black hole the horizon is not a special place. Components of the curvature tensor remain small. In addition the event horizon is a teleological concept; it depends on the complete future time evolution and thus its location cannot be determined by a local measurement. It is thus an interesting question whether we can have a theory where semi-classical gravity remains valid inside the black hole, while maintaining the postulates of black hole complementary.

Black hole complementarity as advocated in \cite{Susskind:1993if} takes the point of view that a semi-classical description can be maintained if semi-classical gravity is valid for each observer separately, although there is no global picture which is physical. It was claimed that no experiment exists that a single observer (possibly utilizing helpers) can perform that will yield a contradiction with the standard rules of quantum mechanics and gravity. Different observers have a \emph{complementary} view of the degrees of freedom which constitute Hawking radiation. After the Page time an observer outside the black hole would find the information about the black hole interior in the radiation, while for an infalling observer the information remains inside the black hole. However, it was convincingly argued in \cite{Mathur:2009hf, Almheiri:2012rt} that the requirement that semi-classical gravity is valid around the horizon\footnote{This was promoted to a fourth postulate in \cite{Almheiri:2012rt}.} is incompatible with the postulates of BHC, even for a single observer.

Below, we will argue that double-holography suggests the following realization of complementarity: there are two complementary descriptions of a black hole which are non-locally related. Information is never copied, but might be encoded in different degrees of freedom in either description. The first, truly quantum gravitational formulation describes only the outside of a black hole geometrically. At or near the horizon we have some unitary quantum mechanical system (we could parametrize our ignorance about this system by a stretched horizon \cite{Susskind:1993if}, or take a concrete model, such as Fuzzballs \cite{Mathur:2005zp}) which produces Hawking radiation. This description obeys the above postulates but violates the equivalence principle, since the quantum mechanical system introduces structure near the black hole horizon. After the Page time, the black hole interior is encoded in the Hawking radiation. The second, approximate description is the usual semi-classical description which describes the geometry past the horizon and respects the equivalence principle. If we do not allow operations which require fine-grained control, both descriptions agree for some finite time outside the black hole. The semi-classical description is a good one, if we only consider simple operations and questions, and it is useful to describe the fate of an infalling observer. Complex operations and questions which depend on fine-grained details (e.g., finding out whether Hawking radiation is pure) can only be faithfully described in the true quantum gravitational picture and are outside the regime of validity of the semi-classical description.\footnote{Morally, this is similar to the proposal for complementarity for computationally bound observers in \cite{Harlow:2013tf}.}

\subsection{Double Holography as a Model for BHC?}
Based on the previous sections, we now want to speculate about a realization of BHC while at the same time resolving the tension between the semi-classical description of black holes and unitarity. Our speculation is based on taking the concepts of double holography discussed above and extrapolating them to the language of general black hole physics. This strategy has an important advantage: Instead of relying on physical intuition for guessing which properties a semi-classical approximation to quantum gravity might have, we can take double holography as a model and study its properties. Carefully extrapolating features of the above model to general black holes then gives a consistent\footnote{Of course, this is only consistent to the extent to which the doubly-holographic models are consistent, as well as to the extent to which the extrapolation is done correctly.} mechanism for BHC. Whether the result of this procedure is indeed the way nature chooses to operate can of course not be decided this way, but at least it will yield a mechanism that seems to be realizable within string theory.

The central idea of our speculation is that the mechanism behind double holography holds in any system with a black hole, if we replace \emph{boundary perspective} by \emph{quantum gravitational description} as well as replace \emph{brane perspective} by \emph{semi-classical description}. We will now sketch some consequences.

A black hole has a quantum gravitational description as a quantum mechanical system with a finite number of degrees of freedom, which is exponential in the black hole entropy. The quantum mechanical system that describes the black hole can be thought of as living near where the horizon would be in a semi-classical description. This quantum mechanical system plays the same role as the degrees of freedom localized to the boundary of the BCFT. It is a system with a large number of degrees of freedom which is coupled to the ``exterior''. It absorbs signals from the outside region, scrambles them and emits them back. In particular the full system is unitary. Of course, black hole complementarity requires that semi-classical gravity is valid outside the stretched horizon, while in double holography the bath region is non-gravitational. However, we see no issue with defining the ambient CFT in the boundary perspective on a dynamical background, since all energies are small and no horizons appear in the bath. We stress that we explicitly do not claim that the exterior description of the black hole is semi-classical gravity outside the black hole, where we have simply traced over the interior degrees of freedom. Rather, we need to introduce some (as of yet unknown) chaotic system at the stretched horizon which can absorb and emit quanta. As the black hole evaporates the Hawking quanta encode more and more information about the black hole microstate.

In order to talk about the interior of the black hole, we need to resort to the semi-classical description. The semi-classical description will be a reorganization of quantum gravity degrees of freedom, the same way that the brane perspective in double holography is a reorganization of the boundary perspective. This reorganization is trivial for simple operators outside the horizon, but highly non-trivial and non-local for complex quantities. The most striking example of this is that even after the Page time, in the semi-classical description the Hawking radiation does not encode the interior of the black hole. Rather, the Hawking quanta are in a thermal state and the interior of the black hole is encoded in local QFT degrees of freedom inside the horizon. This view has important consequences In all descriptions, information about the black hole is only in one place. The information about the black hole interior in the outside description has been traded for parts of the interior of the black hole in the semi-classical description; In other words, given that Hawkings calculation was done in the semi-classical picture it is completely correct. Consequently, for an outside observer in the semi-classical description, unitarity fails. Within the semi-classical description, this does not pose a problem, since---like the brane perspective---we only assume it to be valid for sufficiently simple operations and testing whether a state on $e^S$ qubits is pure is not simple. The question whether or not time-evolution is unitary for an outside observer is therefore outside the regime of validity of the semi-classical description and we should not expect it to give the correct answer.

More generally, for simple outside observers, we can (at least for a finite time) go seamlessly between the quantum gravitational and semi-classical description. Only if we have an observer that can perform complicated operations, e.g., decoding Hawking radiation, is the semi-classical description not longer valid. For instance, consider a powerful outside observer that applies a well-chosen unitary to the radiation of a sufficiently old black hole. Such a (local) interaction has to be described within the quantum gravitational description, since it lies outside the regime of validity of the semi-classical picture. We might still insist on using the map between the two descriptions to see what such an operation would do to the semi-classical description. One can choose the unitary operator such that its effect in the semi-classical picture is to change the state of the quantum fields inside the black hole. Thus, if we insist on describing complex operations in the semi-classical picture, they can look like non-local interactions. While non-local effects are usually not included in semi-classical gravity, this suggests that an agent which can perform (some) complex operations can be \emph{modelled} in semi-classical language by introducing non-local terms into the Lagrangian.

It might be surprising that we are not allowed to ask complex low-energy questions in the semi-classical description, but we do not think that this is an unacceptable consequence. And in fact, the idea that computational complexity or the complexity of operators plays an important role in the validity of semi-classical gravity is a recurring theme \cite{Harlow:2013tf, Kim:2020cds}. More generally, that the validity of some theory breaks down depending on which processes are allowed is very standard. Newtonian physics is only valid as long as all velocities are much smaller than the speed of light, and thermodynamics is only valid as long as we have no experimenter which can control a significant number of molecules in our gas to high precision. For example, if they could arrange for all gas molecules in a box to move to one side, the thermodynamical description breaks down. The obvious difference between this case and black hole physics is that the powerful experimenter needs access to the radiation in the gas to rearrange the molecules, while for black holes access to radiation far away from the horizon is sufficient. From the point of view of a semi-classical observer, the complex observer can seemingly act \emph{non-locally}. However, in the quantum gravitational description the complex observer acts according to the standard rules of local quantum mechanics. It has been clear since \cite{Mathur:2009hf, Almheiri:2012rt} that at least one of the assumptions of BHC needs to be violated. In a model based on double holography two assumptions fail: the equivalence principle in the quantum gravitational description and (at least to describe certain processes) locality in the semi-classical description.

\subsection{Comment on Firewalls and ER=EPR}
We will end with a brief comment on the paper by AMPS \cite{Almheiri:2012rt}, see also Mathur\cite{Mathur:2009hf}, and argue that the resolution to the firewall paradox which follows from our comments agrees with that of ER=EPR \cite{Maldacena:2013xja}.

The authors of \cite{Almheiri:2012rt} posed the following paradox: consider an evaporating black hole. In the semi-classical description and from the point of view of an infalling observer, the horizon is approximately in vacuum. This implies that modes on both sides of the horizon are entangled. Thus, a Hawking mode which escapes the black hole is entangled with a partner mode behind the the horizon. Simultaneously, assuming unitarity of black hole evaporation, after the Page time the escaping radiation is purified by the early Hawking radiation. Thus, the escaping mode must be entangled with a partner mode behind the horizon as well as the early Hawking radiation. This violates monogamy of entanglement. 

Moreover, this violation of monogamy of entanglement can be checked experimentally in the frame of a single observer\footnote{Under certain assumptions on the speed with which computations can be done, it has been argued in \cite{Harlow:2013tf} that such an experiment is not possible. We will not make any assumption about how fast certain computations can be done.}. A sufficiently powerful observer with knowledge of the rules of quantum gravity could collect all Hawking radiation past the Page time and extract a mode entangled with a particular late Hawking quantum. They can then jump into the black hole and check whether the corresponding late mode is entangled with its partner mode behind the horizon, or the mode extracted by the observer. AMPS argued that since the observer acted far away from the black hole, their action cannot possibly have changed anything near the horizon and if black hole evaporation is unitary, the observer must find that their extracted mode is in fact entangled with the near horizon mode. Thus, in order to ensure that early Hawking radiation is purified by late radiation, there must be a mechanism which removes entanglement between late Hawking radiation and their infalling partner modes. This would create a firewall just behind the horizon.

We can reanalyze this gedankenexperiment using the our insights from double holography. In our case there is no single description of an old black hole which obeys both, the equivalence principle as well as unitary time evolution. If we imagine an observer to be powerful enough to extract the information from the radiation, the semi-classical description is not appropriate. In order to describe them extracting the interior partner from the Hawking radiation, we need to use the quantum gravitational description. In order to model the experience of the infalling observer, i.e. the one which would hit the firewall, we need to go to the semi-classical picture. However, since the mapping between those two perspectives is \emph{non-local} it is clear that observer acting on the radiation in the quantum gravitational picture can affect the black hole interior. In particular, the mode extracted by them in the quantum gravitational formulation is the same mode the late quantum is entangled with in the semi-classical perspective. In extracting this mode, the observer creates an excitation behind the horizon. This idea has been dubbed $A = R_b$ \cite{Susskind:2013tg}.

It has been argued in AMPSS \cite{Almheiri:2013hfa} that the observer could simply measure a few particles in the Hawking radiation and jump into the black hole. If the mode behind the horizon was encoded in the radiation, measuring the radiation should disturb the mode and create a firewall. However, it is clear from our doubly-holographic model that the interior of the Black hole is encoded robustly in the radiation. 
The counterargument in AMPSS why modes in the radiation cannot be identified with modes in the interior or the black hole is indeed circumvented by usual quantum error correction \cite{Almheiri:2014lwa}. If we exclude a small region from $A$ in \cref{fig:connected_RT_surface} the region of the associated entanglement wedge near the brane will not change. Thus, if we make a local measurement we can still access the black hole interior from the remaining modes of the radiation which have not been measured.

We should point out that this resolution is not new at all. The arguments given above are the same arguments given in as part of the ER = EPR proposal. However, the important point here is that acting on the radiation happens in a different representation of the quantum state than falling through the horizon. Going between the descriptions is the non-local ``mechanism'' by which the observer creates an excitation behind the horizon.

\section{Conclusions}
\label{sec:discussion}
Assuming entanglement wedge reconstruction, we have argued that in doubly-holographic scenarios the island formula in the brane perspective computes the von Neumann entropy of a subregion $A$ in a different description of the system---the boundary perspective. The result does not need to agree with the von Neumann entropy of $A$ in the brane perspective. One example is a doubly-holographic, eternal black hole coupled to a bath, where the von Neumann entropy of the radiation in the brane perspective grows without bound, while the von Neumann entropy in the boundary perspective ($=$ island entropy in the brane perspective) follows the Page curve. Our observations resonate nicely with older and recent ideas in the literature such as the importance of distinguishing simple and complex quantities \cite{Brown:2019rox,Engelhardt:2021mue}, the possible breakdown of semi-classical gravity when complex operations are allowed \cite{Harlow:2013tf, Maldacena:2013xja, Kim:2020cds}, as well as the possibility of structure at the black hole horizon \cite{Mathur:2005zp,Giveon:2019twx,Giveon:2019gfk,Nomura:2020ska, Langhoff:2020jqa, Matsuo:2020ypv, Hayden:2020vyo, Jafferis:2021ywg}. Going between the boundary perspective and the brane perspective means reorganising the quantum state into different degrees of freedom. Note that this interpretation does not rely on ensemble averaging. However, it does rely on the fact that the reorganization is sufficiently complex. As we have seen, this is indeed the case in double holography. One might hope that the lessons from double holography carry over to ``real life'' quantum gravity, such that we can use it as a very concrete model to extract lessons for quantum gravity in more general contexts.

If the island formula in any semi-classical theory of gravity works analogously to how it works in double holography, it answers the question: ``Given a semi-classical description of a black hole, what is the entropy of Hawking radiation contained in a subregion $A$, if it was computed in a quantum gravitational description?'' This quantum gravitational description does not need to obey the equivalence principle, and in fact, we suggest that it does not. Moreover, the Hawking radiation in the semi-classical description would not contain any information about the black hole interior.

One can generalize the lessons learned from the island formula and try to understand how other quantities get non-locally translated between the brane and boundary description. We have done so for generic complex and simple operators. While we found that simple operators in the bath trivially map between the two prescriptions, simple operators inside the black hole in the brane description can map to complex operators in the boundary description, now however in the bath, i.e., outside the black hole. Mapping between the different perspectives introduces a certain non-locality. One example of this non-locality is exhibited by the island formula itself. We need to know the global quantum state as well as the global geometry in the brane perspective to correctly apply it. Knowledge of the reduced density matrix and geometry in region $A$ is not sufficient. A second example of the non-locality is given by an observer in the boundary perspective who has collected all Hawking quanta for more than the Page time of the eternal black hole. Assume that she has fed those quanta into her quantum computer and has extracted a single qubit which corresponds to a mode in the island to which she has coupled a simple apparatus. Of course such an operation could only be described in the boundary perspective, where the black hole interior is fundamentally absent. If we insist on interpreting the action of such an operation from the brane perspective, this situation looks like a non-local coupling between the observer outside the black hole and a mode inside the black hole.

We can moreover ask about the relation between the brane and boundary perspective. Outside the black hole, the brane approximation is a good thermodynamic description of the boundary perspective, in the sense that the boundary description agrees with the brane description for all simple observations. Importantly, information about the black hole interior is simply never present in the radiation in the brane perspective.\footnote{If the relation between the brane and boundary perspective is a good model for the general relation between semi-classical and quantum gravity, this would imply that in the semi-classical picture information about the interior of an evaporating black hole is not present in the radiation, even after the Page time. Instead, there exists a dual description, analogously to the boundary perspective, in which the information is stored in the radiation. After the Page time, only this latter description is valid for fine-grained questions.} 
Inside the black hole, however, the situation is different. After the Page time, the semi-classical interior of the black hole is a complicated reorganization of complex information in the boundary, i.e., quantum gravitational description, which is not accessible to simple measurements. And there is another remarkable aspect: A simple observer can still answer questions using the semi-classical theory, which one would think are only accessible to an observer capable of doing complex measurements. For example, by using the island rule on a region outside the black hole, they can compute the entropy contained in this region if a very complex observer was to do this computation. It would certainly be interesting to understand which complex quantities can be determined by a simple observer through some non-local law, or more generally to which extent we can describe complex operations on the radiation by additional non-local physics in the brane formulation. 

One of the many puzzling features of the island formula is that whether or not one should include it in semi-classical gravity seems somewhat arbitrary. In fact, it has been argued in \cite{Harlow:2020bee} that JT gravity can be quantized without problem without topology changing processes and also thus without the island formula. Our results suggest a fairly obvious solution to the question whether the island formula should or should not be used. From our point of view, semi-classical gravity might well exist as an (effective) quantum field theory independently of whether and how it is UV completed. The island formula in fact makes no direct statement about states in semi-classical gravity. However, if semi-classical gravity has a UV completion with properties similar to doubly-holographic models (e.g., the existence of different descriptions which are non-locally related) the island formula can be used to compute von Neumann entropies in this quantum gravitational formulation. The fact that the island formula can be derived from semi-classical gravity perhaps indicates that these are the only consistent UV completions in which black holes have a finite Bekenstein-Hawking entropy.

There is a related puzzle. To derive the island formula from semi-classical gravity, we need to allow for Euclidean wormholes. This gives rise to the factorization problem. In \cref{sec:model} we have used a toy model to argue that the island formula as well as wormholes occur in semi-classical theories, because we want to emulate the computation of a quantity of the fully quantum gravitational theory from the semi-classical point of view. This suggest that once one uses the correct UV complete description the factorization problem should not be present. 

The above results still leave many questions unanswered. We have merely suggested that the existence of a representation of a black hole as a quantum mechanical object---``a piece of coal''---can be made consistent with a semi-classical description that has a smooth horizon. However, we have not answered the question what this quantum gravitational formulation could be. Options include Fuzzballs, stretched horizons or firewalls. We would like to point out that structure at the horizon of black holes was very recently discussed in \cite{Hayden:2020vyo}. There, the existence of structure at the horizon was one possible solution to resolve a tension between two statements in quantum information and black hole thermodynamics. It should be interesting to understand if the duality between the brane and boundary perspective helps relieve this tension. In order to completely carry over the lessons from double holography to the general black hole picture it would moreover be necessary to generalize the definition of simple and complex observables from holographic CFTs to quantum field theory. 

Our arguments also call the reliability of semi-classical time evolution into question. It is clear from the above that semi-classical time evolution can only possibly be correct as long as we forbid action with complex operators. For example, we could imagine how in the boundary perspective, Alice collects quanta of radiation of a black hole described by the defect degrees of freedom in which Bob has dropped a note. From this point of view it is clear that once she has collected more than half the total radiation qubits, she can start extracting information about the black hole and thus the note. However, from the brane point of view the radiation does not contain any information about the black hole interior. Collecting radiation in the brane perspective will never give Alice access to Bobs note. If we allow Alice to collect and act on the radiation instantaneously, semi-classical time evolution would not be valid for any finite time. One obvious way of solving this problem is to restrict to observers which can only do simple measurements. However, we should expect a finite time scale even in this case. The most obvious one is the Poincar\'e recurrence time. If we want to grant Alice the power of performing complex operations the validity of semi-classical time evolution most likely depends on bounds of information processing, such as the ones discussed in \cite{Harlow:2013tf, Kim:2020cds}.

\subsection*{Acknowledgements}
I would like to thank Netta Engelhardt, Samir Mathur and particularly Rob Myers for interesting discussions as well as comments on the draft. This work has benefited from discussions and collaborations with Vincent Chen, Ji-Hoon Lee, Rob Myers, Ignacio Reyes, Joshua Sandor, and Ashish Shukla. I acknowledge support by the Simons Foundation through the ``It from Qubit'' collaboration. Research at Perimeter Institute is supported in part by the Government of Canada through the Department of Innovation, Science and Economic Development Canada and by the Province of Ontario through the Ministry of Colleges and Universities.

\bibliographystyle{utphys}
\bibliography{references}

\end{document}